\documentclass{aastex63}

\usepackage{hyperref}
\usepackage{amsmath}
\usepackage{mathtools}
\usepackage{mathrsfs}
\usepackage{amsbsy}
\usepackage{graphicx}
\usepackage{epstopdf}

\newcommand{\sn}[2]{{#1} \times 10^{#2}}

\newcommand{\pd}[2]{\frac{\partial {#1}}{\partial {#2}}}
\newcommand{\pdd}[2]{\frac{\partial^2 {#1}}{\partial {#2}^2}}

\newcommand{\vect}[1]{\textbf{#1}}
\newcommand{\divr}[1]{\nabla_r \cdot {#1}}
\newcommand{\divp}[1]{\nabla_p \cdot {#1}}
\newcommand{\fa}{f_a}
\newcommand{\jab}{\vect{j}^{a/b}}
\newcommand{\phib}{\Phi_b}
\newcommand{\psib}{\Psi_b}
\newcommand{\dvp}{\textrm{d}^3\vect{p}}
\newcommand{\mathcalbf}[1]{\pmb{\mathcal{#1}}}
\DeclarePairedDelimiter{\norm}{\lVert}{\rVert}
\DeclareMathOperator\erf{erf}
\accepted{August 21, 2020}

\usepackage[normalem]{ulem}

\shorttitle{Kinetic Particle Transport in Flares}
\shortauthors{Allred et al.}

\begin{document}

\title{Modeling the Transport of Nonthermal Particles in Flares Using Fokker-Planck Kinetic Theory}

\correspondingauthor{Joel C. Allred}
\email{joel.c.allred@nasa.gov}

\author[0000-0003-4227-6809]{Joel C. Allred}
\affiliation{NASA Goddard Space Flight Center, Solar Physics Laboratory, Code 671, Greenbelt, MD 20771, USA}

\author[0000-0003-2932-3623]{Meriem Alaoui}
\affiliation{Catholic University of America at NASA Goddard Space Flight Center, Solar Physics Laboratory, Code 671, Greenbelt, MD 20771, USA}

\author[0000-0001-7458-1176]{Adam F. Kowalski}
\affiliation{National Solar Observatory, University of Colorado Boulder, 3665 Discovery Drive, Boulder, CO 80303, USA}
\affiliation{Department of Astrophysical and Planetary Sciences, University of Colorado, Boulder, 2000 Colorado Ave, CO 80305, USA}
\affiliation{Laboratory for Atmospheric and Space Physics, University of Colorado Boulder, 3665 Discovery Drive, Boulder, CO 80303, USA.}

\author[0000-0001-5316-914X]{Graham S. Kerr}
\affiliation{Catholic University of America at NASA Goddard Space Flight Center, Solar Physics Laboratory, Code 671, Greenbelt, MD 20771, USA}

\begin{abstract}
We describe a new approach for modeling the transport of high energy particles accelerated during flares from the acceleration region in the solar corona until their eventual thermalization in the flare footpoint. Our technique numerically solves the Fokker-Planck equation and includes forces corresponding to Coulomb collisions in a flux loop with nonuniform ionization, synchrotron emission reaction, magnetic mirroring and a return current electric field. Our solution to the Fokker-Planck equation includes second-order pitch angle and momentum diffusion. It is applicable to particles of arbitrary mass and charge. By tracking the collisions, we predict the bremsstrahlung produced as these particles interact with the ambient stellar atmosphere. This can be compared directly with observations and used to constrain the accelerated particle energy distribution. We have named our numerical code FP and have distributed it for general use. We demonstrate its effectiveness in several test cases.
\end{abstract}


\section{Introduction} \label{sec:intro}
During solar and stellar flares numerous particles are accelerated to high energies. This acceleration likely results from the release of energy during the reconnection of magnetic field occurring near the tops of magnetic flux loops in the corona. These high energy particles propagate along magnetic field lines toward the lower atmosphere. We detect their presence through bursts of nonthermal radiation produced by their interactions with the ambient atmosphere during their transport. The bulk of this emission is produced when the particles reach sufficient density, typically in the footpoints, to slow them to thermal speeds. Details of these particles' energy distributions can be determined by analyzing the spectra of these radiation bursts either through a forward fit of an electron model or inversion of the X-ray spectrum. For example, \citet{2003ApJ...595L..97H} used X-rays observed by the Ramaty High Energy Solar Spectroscopic Imager \citep[RHESSI;][]{2002SoPh..210....3L} of a powerful solar flare to determine the energy spectrum of nonthermal electrons in the flare footpoints. Similar X-ray spectral analyses have been performed in numerous studies \citep[e.g.,][and references therein]{2011SSRv..159..301K, 2012ApJ...759...71E, 2013A&A...551A.135S, 2016ApJ...832...27A, 2019SoPh..294..105A}. 

Forces acting during the particles' transport from looptop to footpoint can significantly alter their energy distributions. If the measured distributions are to be compared to predictions from acceleration models \citep[see e.g.,][and references therein]{2002SSRv..101....1A}, the effects of these forces must be taken into account. Much progress has been made in constructing models of nonthermal particle transport in the presence of strong forces. A kinetic Fokker-Planck description of the distribution function evolution for particles experiencing Coulomb collisions and other forces has been developed by many authors \citep{1943RvMP...15....1C, 1949PhRv...76..220S, 1957PhRv..107....1R, 1962pfig.book.....S, 1965RvPP....1..105T}. In the context of solar flares, \citet{1981ApJ...251..781L} solved a Fokker-Planck equation using a finite difference method for nonthermal electrons including cold-target Coulomb collisions and mirroring due to converging magnetic fields. Here the cold-target assumption means that the speed of the nonthermal electrons greatly exceeds the average speed of thermal particles. \citet{1990ApJ...359..524M} improved upon \citet{1981ApJ...251..781L} by accounting for relativistic effects and energy lost due to synchrotron emission. \citet{1997ApJ...483..496M} and \citet{2019ApJ...880..136J} further improved on \citet{1981ApJ...251..781L} by including the momentum diffusion term ($\vect{D}$ in our notation; discussed in \S~\ref{sec:coulcoll}) which is important for modeling the thermalization of lower energy particles. \citet{2012ApJ...752....4B} used a stochastic differential equation approach to model the source sizes of X-rays produced by nonthermal electrons injected at the top of loop models. \citet{2018ApJ...862..158E} used a collisional transport model to consider the effects of momentum and pitch-angle diffusion on the energy deposition rate from nonthermal electrons.

In addition to electrons, protons and heavier ions are likely accelerated during flares \citep[e.g.,][]{2006ApJ...644L..93H}. \citet{2012ApJ...759...71E} found the energy in ions to be comparable to that of electrons in several large flares. \citet{1986ApJ...309..409T} studied energy and momentum deposition in flare loops produced by beams of nonthermal protons. They considered cold-target and warm-target cases. Here warm-target means that the nonthermal protons had velocities between those of the thermal electrons and ions. They found that in the warm-target approximation, protons were able to penetrate to higher column depths than expected in the cold-target approximation. \citet{1998ApJ...498..441E} considered the hydrodynamic response of flare loops heated by beams of nonthermal protons. \citet{2005AdSpR..35.1743G} compared the hydrodynamic response of loops heated by proton beam to those heated by electron beams. However, the lack of observational constraints on the accelerated ion distribution has precluded their widespread use in flare modelling, despite their likely presence and importance.

During flares the flux of nonthermal particles injected at looptops can be very large. For example, \citet{2003ApJ...595L..97H} found an injection rate greater than $10^{36}$ electrons $s^{-1}$ during an X-class flare. This corresponds to a very large electric current, which if unbalanced, would produce a large magnetic field that would choke off the the propagation of nonthermal electrons away from acceleration site. In order to maintain a steady beam of nonthermal particles (flare observations show that particles are accelerated into the lower atmosphere over time scales much longer than the transient electrodynamical and time-of-flight time scales) and to maintain charge neutrality a co-spatial counter-streaming return current is required. It is driven by an electric field induced in the plasma by the streaming nonthermal particles \citep{1990A&A...234..496V}. This electric field accelerates ambient plasma electrons but acts to slow the downward directed nonthermal particles. In fact, this force often dominates over much of the distance traveled by the nonthermal particles. Therefore, it \emph{must} be included in determining the evolution of the nonthermal particle distribution function.

Considerable progress has been made in developing a theory of return current. \citet{1977ApJ...218..306K} first investigated a nonthermal beam/return current system in one dimension and deduced that the return current can result in additional heating in the upper atmosphere. This was improved by \citet{1980ApJ...235.1055E} to include collisional deceleration. \citet{2005A&A...432.1033Z, 2006ApJ...651..553Z} performed numerical simulations of the effects of return current electric force on nonthermal electron beam propagation and calculated the corresponding hard X-ray spectra. These simulations tracked the hydrodynamic evolution of a flux loop to heating from the nonthermal beam as well as Joule heating produced from the return current. \citet{2012ApJ...745...52H} studied signatures of the return current on the hard X-ray bremsstrahlung emitted during large flares and found the return current can cause a flattening in X-ray spectra at low-energy. \citet{2017ApJ...851...78A} used the Return Current Collisional Thick Target Model (RCCTTM) developed by \citet{2012ApJ...745...52H} and analyzed flattenings in the spectra of several large flares. Under the assumption that these were caused by the return current electric field, they used RCCTTM to fit the observed hard X-ray flux to deduce the injected electron flux density. RCCTTM separates the effects of return current from Coulomb collisions. The return current is assumed to dominate (and collisional deceleration is neglected) while nonthermal electrons propagate through the relatively low density corona. Conversely, in the footpoint collisions are assumed to dominate and the return current electric force is neglected. This was found to work well for several flares, but in others it was found that the coronal density is sufficiently high that collisions cannot be neglected in the corona and the RCCTTM assumptions break down. The authors further concluded that if the injected low-energy cutoff is $\gtrsim$ 60 keV (flare and time dependent) the RC is stable to current-driven instabilities, but if $E_c\sim\delta$ kT or the area of the loop is smaller than that observed with RHESSI, current-driven instabilities can arise. \citet{2008A&A...487..337B} found that the difference between the looptop and footpoint X-ray spectra show a difference higher than can be explained by the classical cold target model \citep{1971SoPh...18..489B}, and attributed this difference to RC losses in the corona. They further concluded that the resistivity is enhanced and current-driven instabilities could be responsible.

In this work, we present a more comprehensive computational model, FP, that solves the Fokker-Planck equation for the transport of nonthermal particles through magnetic flux loops and including effects of strong forces on their energy distributions. \citet{1991A&A...251..693M} demonstrated that the Fokker-Planck equation is equivalent to a coupled set of stochastic ordinary differential equations. \citet{1996ApJS..103..255P} compared the stochastic method with the finite difference method and found that the stochastic method was computationally more expensive. Therefore, in this work we employ a finite difference method. We have designed this method to be general so that it can model many different types of nonthermal particles (e.g., electrons, protons, and ions) as well as general ambient atmospheric conditions. FP improves upon previous work in several important ways:

\begin{enumerate}
 \item The injected nonthermal particles can be of arbitrary mass and charge. Our method applies to injected electrons, protons, alphas, and heavier ions.
 
 \item We do \emph{not} make cold- or warm-target assumptions. Our Coulomb collision operator (\S~\ref{sec:coulcoll}) uses general forms for the Rosenbluth potentials (see Eq.~\ref{eqn:potentials} in \S\ref{sec:coulcoll}) for collisions of nonthermal with thermal particles. These are applicable over large ranges of nonthermal particle energy and ambient plasma temperatures. Note that our own earlier implementation of a Fokker-Planck treatment into a flare numerical code \citep{2015ApJ...809..104A} also made no target temperature assumption. 
 
 \item Our solution to the Fokker-Plank equation includes the return current force (\S~\ref{sec:extforces}), and self-consistently solves the resulting nonlinear differential equation. 

 \item Nonthermal particles are injected into loop models produced using the RADYN flare modeling code \citep{1992ApJ...397L..59C,1995ApJ...440L..29C, 1997ApJ...481..500C, 2005ApJ...630..573A, 2015ApJ...809..104A}. RADYN models in detail the chromospheric radiative transfer and its coupling to non-LTE hydrogen, calcium and helium atomic level populations. Our model uses the ionization state number densities from RADYN in calculating the Coulomb collisional operator for collisions with charged particles (\S~\ref{sec:coulcoll}) and neutral particles (\S~\ref{sec:neutcoll}).

 \item FP can be used as a plug-in to the X-ray data analysis tool, OSPEX (Object Spectral Executive; \url{https://hesperia.
gsfc.nasa.gov/ssw/packages/spex/doc/ospex\_explanation.htm}). OSPEX fits X-ray observations during flares from, for example, RHESSI to a model of X-ray production. Since FP can predict bremsstrahlung resulting from nonthermal particles during their transport, OSPEX+FP can be used to forward fit X-ray observations to obtain constraints on the injected nonthermal electron spectra accounting for the numerous transport effects described in \S~\ref{sec:method}. In a forthcoming study we will demonstrate the use of OSPEX+FP to determine injected electron distributions in a large solar flare. 
 \item FP is designed for general use, and we have released it as an open-source computational tool\footnote{Available for download at \url{https://github.com/solarFP/FP}}. This allows FP to be easily incorporated into numerical models of flares driven by nonthermal particles.
 
\end{enumerate}
As we have discussed previously, one of the main motivations for the development of FP was to more accurately model the transport of nonthermal particles during flares, and the resulting impacts on stellar atmospheres. The flare modeling code RADYN previously included a Fokker-Planck solver based on the work by \citet{1990ApJ...359..524M}. This was described in \citet{2015ApJ...809..104A}. That version was similar to what we describe but we have enhanced it to self-consistently handle the force due to return currents and second order parallel momentum (energy) diffusion. This updated Fokker-Plank treatment is currently being incorporated into RADYN. The predicted heating rates due to nonthermal particles can be very different for models that include return current effects. An example is shown in \S\ref{sec:ccttm}. 

We wish to stress that while FP represents a marked improvement over our prior implementation of the Fokker-Plank treatment in RADYN, that prior effort did include many of the transport effects discussed here. Namely, since \cite{2015ApJ...809..104A} RADYN has made no assumption as target temperature (there is no cold- or warm- target approximation). 

In this paper we will describe our new Fokker-Planck nonthermal particle transport code, FP. In Section~\ref{sec:method} we describe the computational method we employ to solve the Fokker-Planck equation. In Section~\ref{sec:testcases} we demonstrate solutions to several test cases and compare them to results derived analytically. Finally, in Section~\ref{sec:conc} we present our conclusions and discuss plans for future improvements.

\section{Methodology}\label{sec:method}
The semi-relativistic Fokker-Planck kinetic equation for Coulomb collisions and external forces can be written using the notation of \citet{1965RvPP....1..105T} as follows,
\begin{eqnarray}\label{eq:fp}
 \pd{\fa}{t} &+&\divr{\left(\vect{v} \fa \right)} + \divp{ \left( \vect{F}^e \fa \right)}\nonumber\\
 & + & \sum_{b} \divp{\jab} = 0
\end{eqnarray}
where $\fa$ is the distribution function for a particle, $a$, with mass, $m_a$. That is $\fa(\vect{r},\vect{p}, t) \textrm{d}^3\vect{r}\; \dvp\; \textrm{d}t$ is the number of particles of type $a$ within the volume $\vect{r} + \textrm{d}^3\vect{r}$ and with momenta between $\vect{p}$ and $\vect{p} + \dvp$ in the time interval, $t + \textrm{d}t$. $\vect{v}$ is the phase space velocity ($\vect{v} = \vect{p}/\gamma m_{a}$); $\gamma$ is the relativistic Lorentz factor. $\vect{F}^e$ is the sum of external forces and includes all forces except the Coulomb collision force. $\jab/\fa$ is the net force on particles of type $a$ due to Coulomb collisions with particles of type $b$. The sum is over all constituent plasma particles. $(\divr{})$ and $(\divp{})$ are divergence operators in phase space for the geometric and momentum spaces, respectively. In this form, it is evident that Eq.~\ref{eq:fp} is a conservation equation in phase space. 

\subsection{Loop Geometry}
In the low plasma beta environment typical of the solar corona, charged particles are constrained to move along the axis of magnetic field lines. Therefore, we follow the example of many previous works \citep[e.g.,][]{1985ApJ...289..414F, 1999ApJ...521..906A, 2005ApJ...630..573A, 2015ApJ...809..104A, 2009A&A...499..923K, 2009ApJ...702.1553L, 2012A&A...547A..25P, 2015SoPh..290.3487K, 2016ApJ...827...38R} making the assumption that the motion of beams of nonthermal particles traveling from the acceleration region near the top of magnetic loops toward the footpoints is well-characterized by a 1.5D geometry with spatial dimension, $z$, being the distance along a loop measured from the injection site assumed to be at the loop apex and pitch angle, $\theta$, being the angle between a particle's momentum vector and the $z$-axis. We assume that the loop has a constant cross-sectional area. In solving Eq.~\ref{eq:fp}, we assume no variation in directions perpendicular to the magnetic field axis, which implicitly assumes that we obtain $\fa$ for positions near the center of a bundle of magnetic loops each receiving equal injection of nonthermal particles. In the momentum space, we employ a standard spherical coordinate system, where $p$ is the momentum magnitude and $\theta$ and  $\phi$ are the pitch and azimuthal angles, respectively. For this work, we assume azimuthal symmetry, so $\partial / \partial \phi = 0$. We use the momentum space basis vectors, $\hat{p}$, which is a unit vector in the direction of $\vect{p}$, and $\hat{\theta}$, which is perpendicular to $\hat{p}$ and in the direction of increasing $\theta$. For convenience, we introduce $\mu = \cos \theta$. 

The forces in the Fokker-Planck equation are dependent on the ambient plasma density, temperature, and magnetic flux density, so it is important to have a model for how these quantities vary along the loop. As examples, we use the temperature and density stratification from RADYN loop models which have been evolved into a state of hydrodynamic equilibrium. 
The magnetic flux density is from a model that assumes 1000 G in the photosphere and exponentially decreases to 75 G at 3 Mm above the photosphere and is constant above that. While it is true that this magnetic flux density model is ad-hoc, we note that its most important effect, the magnetic mirroring force, is relatively insensitive to inaccuracies in the overall field strength. Often the changing the magnetic flux density is used as a proxy for estimating changes in the loop cross-sectional area, so that magnetic flux is constant within the loop. Since we have assumed a constant loop cross-sectional area we are not making this assumption, but we note that it is straightforward to alter our method to do so. When evaluating the divergence of the geometric fluxes in Eq.~\ref{eq:fp} ($\vect{v} \fa$), a nonuniform area, $A(z)$, which scales with magnetic flux density can be used. 

\begin{figure*}[htb!]
 \includegraphics[scale=0.65]{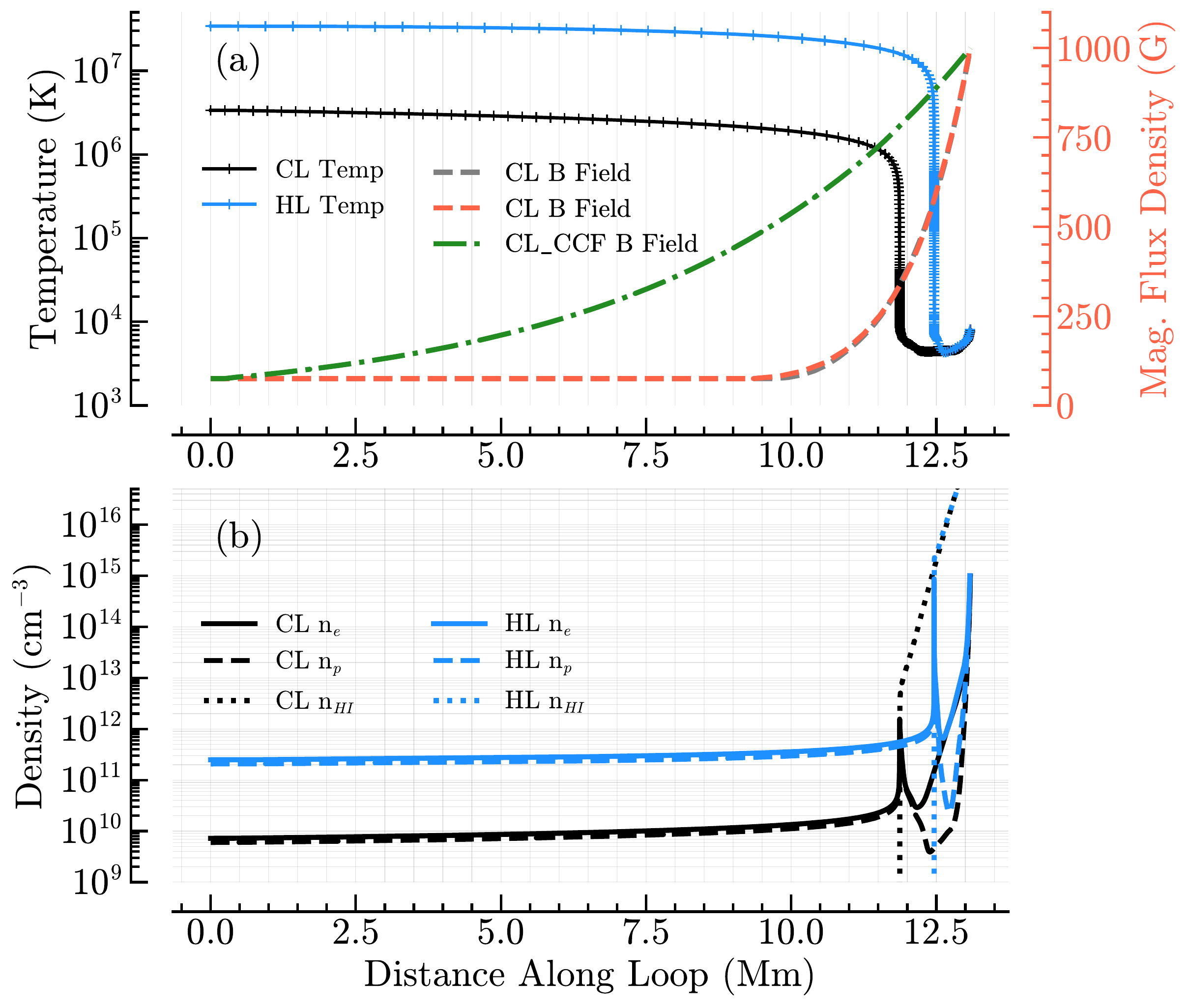}
 \caption{Temperature (top panel solid lines), magnetic flux density (top panel dashed lines), electron density (bottom panel solid lines), proton density (bottom panel dashed lines) and neutral hydrogen density (bottom panel dotted lines) as a function of distance along the loop axis for the CL (black lines) and HL (blue lines) loop models. The CL\_CCF loop model is identical to CL except for its magnetic flux density stratification shown by the dot-dashed green line. Locations of z-axis grid cell centers are indicated by the ``+'' symbols in the temperature plots.} \label{fig:loops}
\end{figure*}

Figure~\ref{fig:loops} shows temperature, density, and magnetic flux density stratification for three loop conditions in hydrodynamic equilibrium obtained from RADYN simulations. Each of these has a half-length, $L$, of 13 Mm, but their densities and temperatures are quite different. One (HL) has an apex temperature of 34 MK which is typical of temperatures measured at the peak of large flares \citep{2014ApJ...781...43C}. The others (CL and CL\_CCF) have a apex temperature of 3.4 MK which is typical of non-flaring active region loops. CL and CL\_CCF are identical except for their magnetic flux density stratifications. CL has uniform magnetic flux density throughout the upper coronal portion of the loop, but CL\_CCF has a strongly converging coronal field.

For fast particles accelerated during flares, the electrodynamic transient and time-of-flight time scales are much less than is observable with current instrumentation. Therefore, for this work we seek a steady state solution consistent with a flux of non-thermal particles injected at the top of flaring loops.

\subsection{Collisions with Charged Particles}\label{sec:coulcoll}
The collision term, $\jab$, can be divided into terms representing dynamic friction ($\vect{F}$) and diffusion ($\vect{D}$) forces. For this work, we consider the collision of nonthermal particles with a thermal plasma. Even in superhot flares ($\sim 50$ MK), thermal particles are mostly non-relativistic and are well-characterized by a Maxwell-Boltzmann distribution with temperature, $T_b$. The collision terms, $\vect{F}$ and $\vect{D}$ can be further divided into terms representing collisions with charged particles ($\vect{F}_c$ and $\vect{D}_c$) and collisions with neutral particles ($\vect{F}_N$ and $\vect{D}_N$). 

Making the assumption that the target particles, $b$, are non-relativistic, the charged particle collision terms can be written as:
\begin{eqnarray} \label{eqn:fd}
 \vect{F}_c &=& -4 \pi K^{a/b} \fa \nabla_p \phib \nonumber\\
 \vect{D}_c &=& 4 \pi K^{a/b} \frac{m_a}{m_b}  \left(\nabla_p \fa \cdot \nabla_p\right) \nabla_p \psib
\end{eqnarray}

Here we make use of the notation that $(\vect{A} \cdot \nabla_p)$ is the momentum space convective operator for a vector, $\vect{A}$. The collection of constants, $K^{a/b}$, is defined as
\begin{equation}
 K^{a/b} = 4 \pi \lambda^{a/b} m_a (e^2 Z_a Z_b)^2
\end{equation}
where $\lambda^{a/b}$ is the Coulomb logarithm, $Z_a$ and $Z_b$ and $m_a$ and $m_b$ are the charges numbers and masses of the particles of type $a$ and $b$, respectively. $e$ is the electron charge in Gaussian units. The expression for $\lambda^{a/b}$ is from Eq.~13 of \citet{2015ApJ...809..104A}, which we reproduce here for clarity. 
\begin{equation}
  \lambda^{a/b} = \ln\left(\frac{M_{ab} v^2}{\hbar} \left[\frac{m_b}{\pi n_b Z_b^2 e^2}\right]^\frac{1}{2}\right)
\end{equation} 
where $M_{ab}$ is the reduced mass of the beam and target particles. $\phib$ and $\psib$ are potential functions derived from the distribution function, $f_b$ for particles of type $b$. They are defined as,
\begin{eqnarray}\label{eqn:potentials}
 \psib(\vect{p}) &=& -\frac{1}{8 \pi} \int |\vect{p} - \vect{p}'| f_b(\vect{p}') \dvp' \nonumber\\
 \phib(\vect{p}) &=& -\frac{1}{4 \pi} \int \frac{f_b(\vect{p}') \dvp'}{|\vect{p} - \vect{p}'|}
\end{eqnarray}
For Maxwell-Boltzmann distributions of the target the potentials have the forms, \citep[Eqs. 17.9 and 17.23 of][]{1965RvPP....1..105T},
\begin{eqnarray}
 \psib  &=& - \frac{n_b}{8 \pi \sqrt{x_b}} \sqrt{2 m_b k T_b} \nonumber\\
             & & \left[ \left(x_b+1\right)\xi'(x_b) + \left(x_b + \frac{1}{2}\right) \xi(x_b)\right]
\end{eqnarray}
and
\begin{equation}
 \phib = - \frac{n_b}{4 \pi m_b v}\left[ \xi'(x_b) + \xi(x_b)\right]
\end{equation}
where 
\begin{eqnarray} \label{eqn:xi}
\xi(x_b) &=& \erf(\sqrt{x_b}) - \xi'(x_b) \nonumber\\
\xi'(x_b) &=& \frac{2}{\sqrt{\pi}}e^{-x_b} \sqrt{x_b} \nonumber\\
x_b &=& \frac{m_b v^2}{ 2 k T_b} = \frac{m_b p^2}{2 m_a^2 k T} 
\end{eqnarray}

In deriving Eqs.~\ref{eqn:fd}-\ref{eqn:xi}, we have not made use of our loop geometry, and they are valid in any 3D geometry. Now imposing azimuthal symmetry and noting that $\phib$ and $\psib$ are functions of only the momentum magnitude greatly simplifies the expressions for $\vect{F}_c$ and $\vect{D}_c$, which can be rewritten as 
\begin{eqnarray}\label{eq:fcdc}
 \vect{F}_c &=& -\frac{m_a}{m_b} \frac{n_b K^{a/b}}{p^2} \xi \fa\; \hat{p} \nonumber\\
 \vect{D}_c &=& -n_b K^{a/b} \frac{\xi}{2 x_b p} \pd{\fa}{p}\; \hat{p} \nonumber\\
          & & -\frac{n_b K^{a/b}}{2 p^2} \left(\xi + \xi' - \frac{\xi}{2 x_b}\right) \pd{\fa}{\theta}\; \hat{\theta}
\end{eqnarray}
$D_p$ is the parallel momentum diffusion term discussed by \citet{2019ApJ...880..136J} and $D_\theta$ is the pitch-angle diffusion term. 

\subsection{Collisions with Neutral Particles} \label{sec:neutcoll}
The stopping force of a neutral gas from collisions with energetic particles is discussed extensively in \citet{evans55} and here we make use of those results. The dynamic friction and diffusion terms can be written as \citep[cf. Eq. 2.24 and 2.25 p. 581][]{evans55, 1949PhRv...76..220S}:
\begin{equation}
 \vect{F}_N = -\frac{m_a}{m_e} \frac{n_N K^{a/N}}{p^2} \fa\; \hat{p}
\end{equation}
\begin{equation}
 \vect{D}_N = -\frac{n_N K'^{a/N} }{2 p^2} \pd{\fa}{\theta}\; \hat{\theta}
\end{equation}
where $K^{a/N}$ and $K'^{a/N}$ are defined as:
\begin{eqnarray}
 K^{a/N} &=& 4 \pi \lambda^{a/N} m_a (e^2 Z_a)^2 Z_N  \nonumber \\
 K'^{a/N} &=&  4 \pi \lambda'^{a/N} m_a (e^2 Z_a Z_N)^2 
\end{eqnarray}
where $Z_N$ is the atomic number of the neutral species and $m_e$ is the electron mass. The Coulomb logarithm, $\lambda^{a/N}$, differs when $a$ represents electrons or ions. They are, 
\begin{eqnarray}
 \lambda^{e/N} &=& \ln\left(\frac{m_e c^2 \beta \gamma \sqrt{\gamma-1} }{I_N}\right) \nonumber \\
 \lambda^{i/N} &=& \ln\left(\frac{2 m_e c^2 \beta^2 \gamma^2}{I_N}\right) 
\end{eqnarray}
for electrons and ions, respectively. Here $\beta$ is the relativistic speed ($\beta = v/c$), and $I_N$ is the ionization energy. The Coulomb logarithm for diffusion from neutral particles is 
\begin{equation}
 \lambda'^{a/N} = \ln\left(\frac{\gamma \beta}{\sqrt{2} Z_N^{\frac{1}{3}} \alpha}\right)
\end{equation}
where $\alpha$ is the fine structure constant.

\subsection{External Forces} \label{sec:extforces}
The external forces considered in this work are the reaction force from synchrotron radiation, $\vect{F}_S$, reflecting due to magnetic mirroring, $\vect{F}_M$, and importantly, the return current force, $\vect{F}_{RC}$. The first two of these are discussed in \citet{1990ApJ...359..524M}, and we simply quote their expressions here,
\begin{eqnarray}
 \vect{F}_S = -S \beta \left(\gamma^2 \left(1-\mu^2\right)\; \hat{p} + \mu \sqrt{1-\mu^2} \; \hat{\theta}\right)
\end{eqnarray}
$S$ is defined by,
\begin{equation}
 S = \frac{2}{3}\frac{\left(Z_a e\right)^4}{m_a^2 c^4} B^2
\end{equation}
and $B$ is the magnetic flux density in gauss. The magnetic mirroring force is given by, 
\begin{equation}
 \vect{F}_M = \frac{1}{2} m_a v^2 \sqrt{1-\mu^2} \frac{\textrm{d} \ln B}{\textrm{d}z} \; \hat{\theta}
\end{equation}

Including the return current force is the main task in this work. We assume that the return current is carried by drifting thermal electrons moving to neutralize the beam current. We neglect instabilities \citep{1977RvGSP..15..113P,2002ASSL..279.....B,2009ApJ...690..189K} and the presence of runaway electrons \citep{1985A&A...142..219R,1985ApJ...293..584H},
therefore, we implicitly assume that the electric field is much less than the Dreicer field \citep{1960PhRv..117..329D}. We make the assumption that, in the steady state, the return current density, $\vect{J}_{RC}$, cancels the beam current density, $\vect{J}_{beam}$, at each point in space. The beam current density is obtained from the distribution function, $\fa$. 
\begin{equation}\label{eqn:jrc}
 \vect{J}_{RC} = - \vect{J}_{beam} = -Z_a e \mathcalbf{F}
\end{equation}
$\mathcalbf{F}$ is the number flux of beam particles. Because of azimuthal symmetry only the field-aligned component has a net contribution to $\mathcalbf{F}$, which is given by, 
\begin{equation}\label{eqn:numflux}
 \mathcalbf{F} = \int \vect{v} \fa \dvp = \int \mu\beta c \fa \dvp \; \hat{z}
\end{equation}

For time scales much longer than the collisional time scale, \citet{1990A&A...234..496V} found Ohm's law to be valid. Since we are considering the steady state, we assume the plasma obeys Ohm's law. That is $\vect{E}_{RC} = \eta \vect{J}_{RC}$, where $\eta$ is the plasma resistivity. The force on beam particles from $\vect{E}_{RC}$ is given by, 
\begin{eqnarray}
 \vect{F}_{RC} = -\left(Z_a e\right)^2 \eta \mathcal{F} \left(\mu \hat{p} - \sqrt{1-\mu^2} \hat{\theta}\right)
\end{eqnarray}
where we have made use of $\hat{z} = \mu\hat{p} - \sqrt{1-\mu^2}\hat{\theta}$. We use an expression for $\eta$ that includes the collisions of drifting ambient electrons with ionized hydrogen \citep{1962pfig.book.....S} and helium \citep{1977PhFl...20..589H} and with neutral hydrogen. The drifting electron/neutral hydrogen collision frequency is from \citet{2012ApJ...753..161M}.

\subsection{Steady State}
We rewrite Eq.~\ref{eq:fp} in a steady state form by setting $\partial \fa/\partial t = 0$, evaluating the divergences, and collecting terms with similar derivatives of $\fa$. After considerable algebraic manipulation we obtain, 
\begin{eqnarray}\label{eqn:ssfp}
&&  A^{(p^2)} \pdd{\fa}{p} + A^{(p)} \pd{\fa}{p}  \nonumber \\ 
&+& A^{(\mu^2)} \pdd{\fa}{\mu} + A^{(\mu)} \pd{\fa}{\mu} \nonumber\\
&+& A^{(z)}\pd{\fa}{z} + A^{(f)} \fa = 0
\end{eqnarray}
where
\begin{subequations}
\begin{equation}
 A^{(p^2)} = -\sum_{b} n_b K^{a/b} \frac{\xi}{2 x_b p}
\end{equation}
\begin{eqnarray}
  A^{(p)} &=& \sum_{b} \left[\frac{n_b K^{a/b}}{p^2} \left( \frac{\xi} {2 x_b} -  \xi' - \frac{m_a}{m_b} \xi \right) \right] \nonumber\\
           &-& \sum_{N}  \left[\frac{m_a}{m_e} \frac{n_N K^{a/N}}{p^2} \right] \nonumber \\
           &-& S \beta \gamma^2 \left(1-\mu^2\right) - \left(Z_a e\right)^2 \eta \mu \mathcal{F} 
\end{eqnarray}
\begin{equation}
  A^{(\mu^2)} = C^{(\mu)} \left(1-\mu^2\right)
\end{equation}  

\begin{eqnarray}
  A^{(\mu)} &=& \frac{S}{\gamma m_a c} \mu\left(1 - \mu^2\right)  -2 \mu C^{(\mu)} \nonumber \\
            &-& \frac{\beta c}{2 \gamma} \left(1-\mu^2\right) \frac{\textrm{d} \ln B}{\textrm{d}z} \nonumber\\
            &-& \frac{\left(Z_a e\right)^2 \eta \mathcal{F} \left(1-\mu^2\right)}{p}
\end{eqnarray}
\begin{eqnarray}
  A^{(z)} &=& \mu \beta c
\end{eqnarray}
\begin{eqnarray}
  A^{(f)} = &-& \sum_{b} \left[\frac{m_a}{m_b} \frac{2 n_b K^{a/b} \xi' x_b}{p^3} \right] \nonumber\\
           &-& \frac{4 S}{\gamma m_a c}\left[\gamma^2 \left(1-\mu^2\right) + \mu^2 -\frac{1}{2}\right] \nonumber\\
           &+& \frac{\mu \beta c}{\gamma} \frac{\textrm{d} \ln B}{\textrm{d}z}
\end{eqnarray}
\begin{eqnarray}
  C^{(\mu)} = &-&
  \sum_{b} \left[\frac{n_b K^{a/b}}{2 p^3} \left(\xi + \xi' - \frac{\xi}{2 x_b}\right)\right] \nonumber\\
           &-& \sum_{N}  \left[\frac{n_N K'^{a/N}}{2 p^3} \right]
\end{eqnarray}          
\end{subequations}
In deriving these expressions, we have neglected terms proportional to $\partial \lambda/\partial p$ because the Coulomb logarithms vary slowly with respect to $p$. 

\subsection{Numerical Method}\label{sec:nummethod}
We solve the time-independent Fokker-Planck equation (Eq.~\ref{eqn:ssfp}) by discretizing the phase space, approximating derivatives as finite differences, and iterating the solution until residuals are below a threshold value. In our field aligned loop approximation there are three phase space dimensions: the $z$-axis ranging from $z = [0,L]$, the $\theta$-axis ranging from $\theta = [0,\pi]$ and the $p$-axis ranging from $p = [p_{min}, p_{max}]$. $p_{min}$ and $p_{max}$ are configurable. It is possible to choose $p_{min} = 0$, but that results in slow convergence to the steady state. A better choice is $p_{min} = \sqrt{2 m_a k T_e}$ where $T_e$ is the ambient electron temperature. This allows the solution to model the thermalization of beam particles while still maintaining good convergence. $p_{max}$ should be chosen to be sufficiently large that the number of particles with momenta above it is negligible. A reasonable choice for flare conditions is $p_{max} = 100 m_a c$. Within these ranges the phase space is divided into grid cells. The $z$-axis grid is provided by the loop model and must have sufficient resolution to resolve temperature and density gradients. For loop models provided by RADYN there are typically 191 grid cells along the $z$-axis. RADYN uses an adaptive grid allowing it to resolve steep gradients in the transition region. The location of z-axis grid cells for the loop models HL and CL are shown using the ``+'' symbol in Fig.~\ref{fig:loops}. The $\theta$- and $p$-axes are divided into evenly and logarithmically spaced grid cells, respectively. The number of cells in each of these dimensions is configurable, but 60 cells in the $\theta$- and 100 cells in $p$-axis are typical numbers based on our numerical experiments. On these grids we construct three point (2nd order accurate) first and second derivative stencils. In solving Eq.~\ref{eqn:ssfp} we employ an upwind scheme so first derivative stencils are one-sided in the upwind direction, except in computing the residual when we use a centered stencil. 

Eq.~\ref{eqn:ssfp} can be represented as a nonlinear matrix equation,
\begin{equation}\label{eqn:mateqn}
 A \fa = b
\end{equation}
$A$ is a function of $\fa$ through the factor, $\mathcal{F}$, in the return current force. The right hand side, $b$, is nonzero only at the boundaries, which are discussed in Section~\ref{sec:bc}.

We solve Eq.~\ref{eqn:mateqn} using an alternating direction implicit iterative scheme. We split the operator, $A$ into components along each axis, that is $A = (A_d + A_p + A_\mu + A_z)$, where $A_d$ contains the diagonal term of $A$ and $A_p$, $A_\mu$ and $A_z$ contain the off-diagonal terms in the $p$, $\mu$ and $z$ directions, respectively. Therefore, we can construct implicit matrix equations in each direction separately. For example, in the $p$-direction Eq.~\ref{eqn:mateqn} is rewritten as $(A_d + A_p) \fa^{n+1} = b - (A_\mu + A_z)\fa^{n}$ where $\fa^{n}$and $\fa^{n+1}$ are the current and next iterative solutions. In each grid cell the direction of the flux is evaluated using $\fa^{n}$ and the correct upwind stencil is chosen. Thus, Eq.~\ref{eqn:mateqn} is approximated as a series of penta-diagonal matrix equations. The penta-diagonal systems are easily solved in sweeps along the $p$-, $\mu$- and $z$-directions. After a complete iteration, $\mathcal{F}$ is reevaluated and the return current components of $A$ are updated. Sweeps are performed until the residual and relative change between $\fa^n$ and $\fa^{n+1}$ are below threshold values. The residual of an equation in the form of Eq.~\ref{eqn:mateqn} is typically defined by $\norm{A \fa -b}/{\norm{b}}$, where $\norm{}$ is the $L^2$-norm. But since $b$ is zero in the interior of our computational domain, we cannot use that. Instead we define a normalized residual, $r$, that measures how closely the components of $A \fa$ sum to zero. We use
\begin{equation}
  r = \frac{\norm{A \fa}}{\norm{ \left(|A_d \fa| + |A_\mu \fa| + |A_p \fa| + |A_z \fa|\right)}}
\end{equation}

\subsection{Boundary Conditions}\label{sec:bc}
Since Eq.~\ref{eqn:mateqn} is homogeneous, the boundary conditions determine the particular solution to which our algorithm converges. Because of azimuthal symmetry, $\partial \fa /\partial \mu = 0$, at the $\mu = \pm 1$ boundaries. At $p = p_{min}$, we assume $\partial \fa /\partial p = 0$. This is applicable, since in the low-energy limit, $\fa$ approaches a thermal distribution. Since $p_{max}$ is chosen to be sufficiently high that the number of particles above it are negligible, we assume $\fa(p_{max})= 0$. At the $z = L$ boundary we have implemented a nonreflecting boundary condition so that particles reaching that depth simply pass through. However, we note that for loop models extending into the subphotosphere the densities are high and very few particles reach the $z = L$ boundary.

The injected particle flux distribution, $\mathcal{F}_0(E,\theta)$, is specified at the $z=0$ boundary. Numerous observations have shown that $\mathcal{F}_0$ is well-represented by a power-law in energy, $E$. So $\mathcal{F}_0$ is written as 
\begin{equation}\label{eq:powerlaw}
\mathcal{F}_0(E,\theta) = \begin{cases}
              \frac{N_0 (\delta-1)}{E_c} M_0(\theta) \left(\frac{E}{E_c}\right)^{-\delta} & \textrm{for $E \ge E_c$} \\
              0 & \textrm{for $E < E_c$}
              \end{cases} 
\end{equation}
where $N_0$ is the total number flux density, $E_c$ is a cutoff energy below which no particles are injected, and $\delta$ is the power-law index. The pitch-angle distribution of the injected particles, $M_0(\theta)$, remains largely unconstrained by current observations. To account for this, FP allows users to select from three general forms for $M_0(\theta)$. Those are fully beamed ($\theta =0$ for all injected particles), isotropic in the forward hemisphere and a Gaussian centered at $\theta = 0$ and with a standard deviation of $\theta_0$.

\subsection{Initial Estimate}
Since Eq.~\ref{eqn:mateqn} is solved iteratively, we must provide an initial estimate of the solution, $\fa^0$. We note that often in flare loops, collisional deceleration dominates other forces. So we make an initial estimate using the analytic solution to a much simpler form of Eq.~\ref{eqn:ssfp} in which $A^{(p^2)} = A^{(\mu^2)} = A^{(\mu)} = A^{(f)} = 0$ and forces in $A^{(p)}$ other than the collision force are neglected. This simple first order differential equation has a solution, $\fa^0(z,\mu,p) = \fa(z=0, \mu, p'(z))$ where $p'(z) = p - \int_0^z A^{(p)} / A^{(z)} dz$. In other words, this solution is just the $z = 0$ boundary condition but shifted by the cumulative effect of collisional decelerations. 

\subsection{Heating rate on the ambient plasma}
Much of the energy lost by the nonthermal particles during their transport is gained as heat by the ambient plasma. FP models this heating and momentum deposition. This is useful for dynamic flare simulations, such as those produced using RADYN. We assume that the energy and momentum lost due to Coulomb collisions and due to the return current is gained by the local ambient plasma. Since synchrotron emission is usually optically thin, we assume that it has no local heating effect. Using the solution to the distribution function the heating rate, $Q$, and the momentum deposition rate, $\vect{G}$, are obtained from
\begin{eqnarray}
  Q = -\int \left(\vect{F}_t \cdot \vect{v}\right) \fa \dvp \\
  \vect{G} = -\int \left(\vect{F}_t \cdot \hat{z}\right) \fa \dvp \; \hat{z}
\end{eqnarray}
where $\vect{F}_t$ is the total of the Coulomb collision and return current forces. 
\section{Test Cases} \label{sec:testcases}
In this section we demonstrate the performance of FP and compare its predictions to previous work using several test cases. 
\subsection{Collisional Cold Thick Target}\label{sec:ccttm}
\begin{figure*}
\includegraphics[scale=.475]{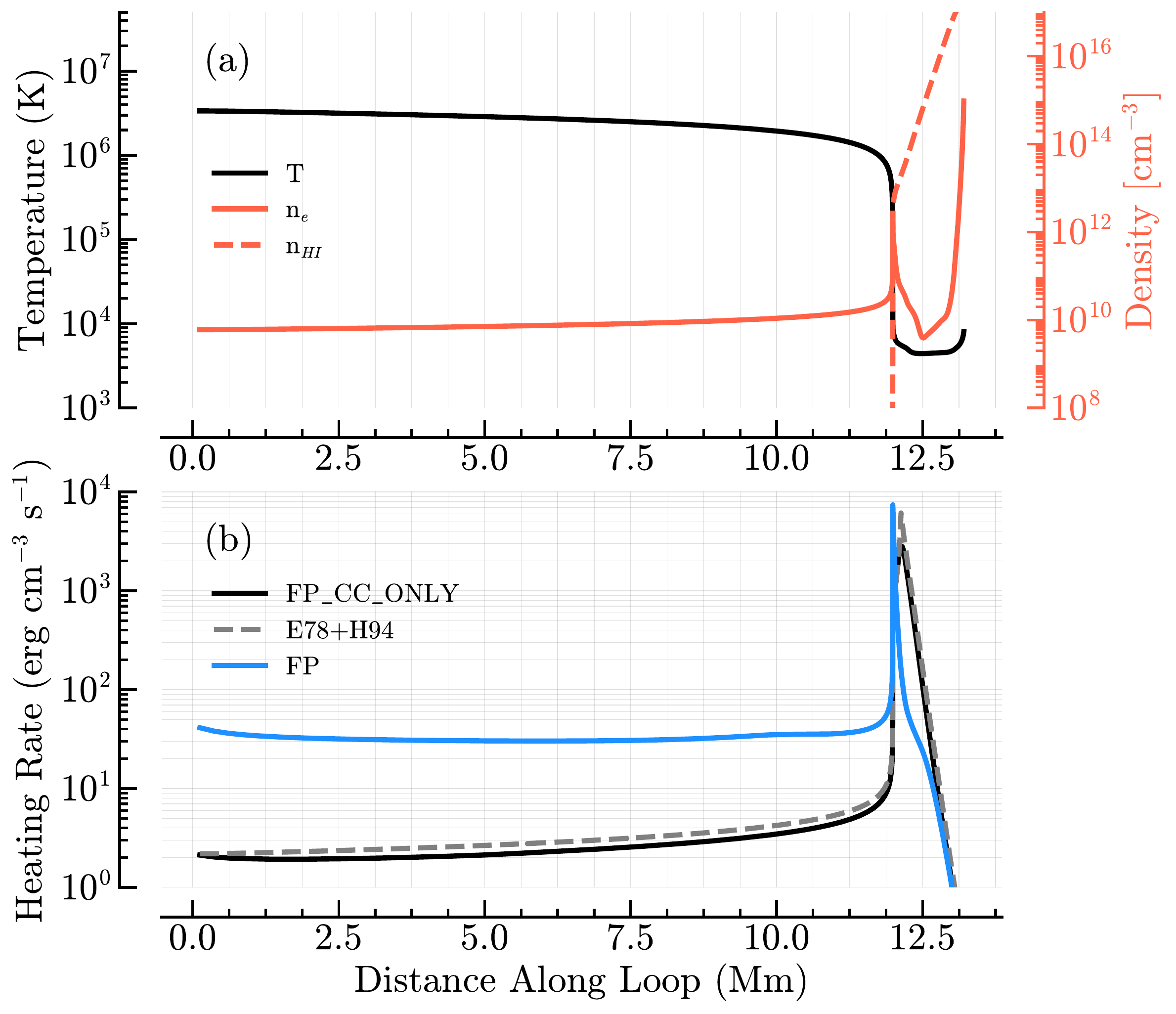}
\includegraphics[scale=.475]{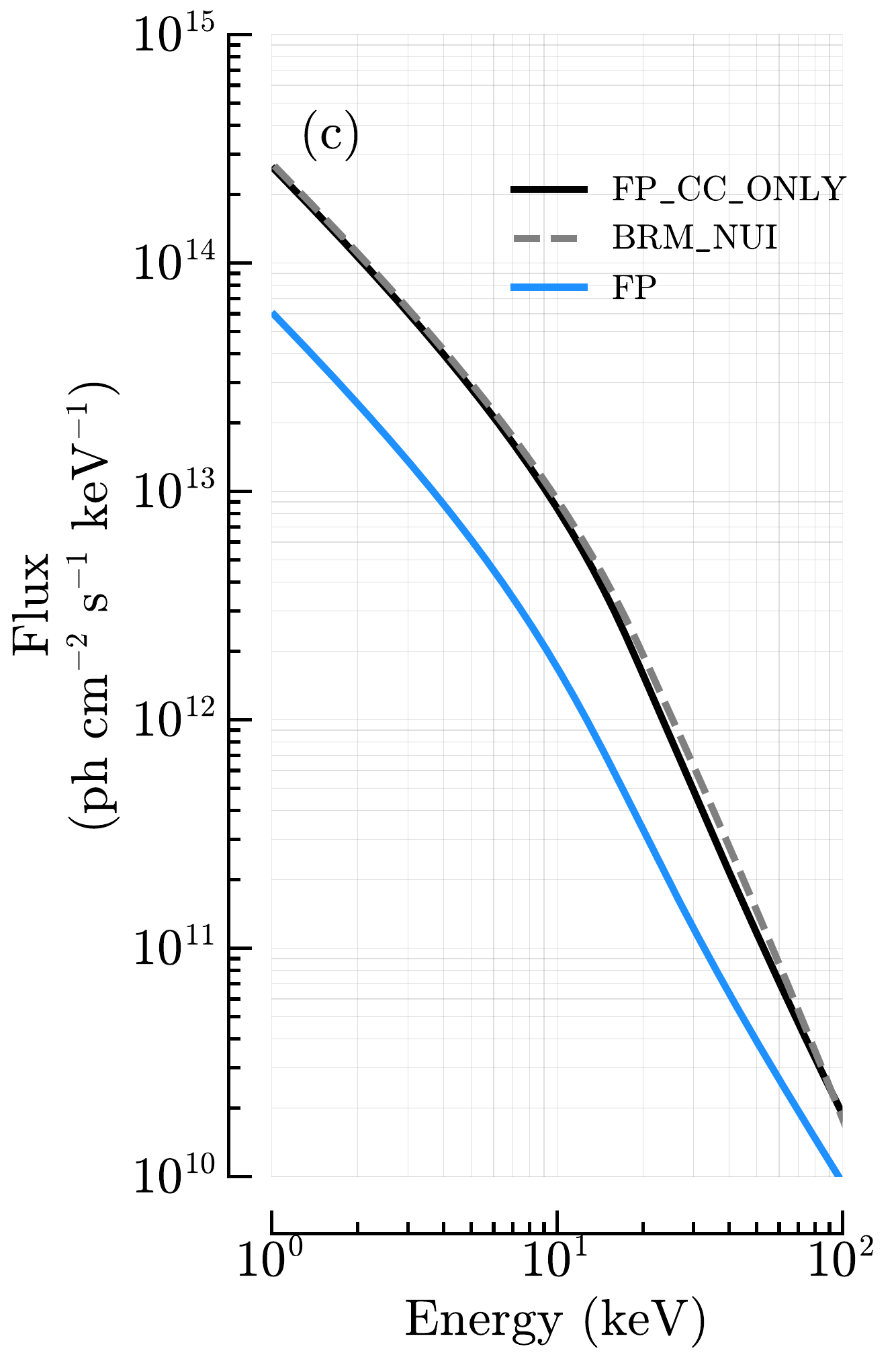}
\caption{(a) The temperature (black line), electron density (red line), and neutral hydrogen density (red dashed line) in a pure hydrogen version of the CL loop model as a function of distance from the loop apex. (b) The volumetric heating rate predicted by FP\_CC\_only (black line), compared to the analytic expressions derived by E78 and HF94 (grey dashed line), and the full FP predicted heating rate (blue line). (c) Spatially integrated bremsstrahlung photon spectra from the FP\_CC\_only (black line), BRM2\_NUI (gray dashed line), and FP (blue line) models.}\label{fig:fpe78}
\end{figure*}
The collisional cold thick target model \citep[CCTTM;][]{1971SoPh...18..489B} has been of fundamental importance to the study of flares. Briefly, the model assumes that fast electrons accelerated in the corona travel toward the chromosphere. The electrons are slowed by Coulomb collisions, but because they are fast compared to the ambient thermal speed, the cold-target approximation is valid. Return currents and other transport forces are ignored by the CCTTM.  Using the CCTTM \citet{1978ApJ...224..241E} (hereafter, E78) derived an analytic model for how nonthermal electrons (or protons) deposit energy into the ambient atmosphere assuming an arbitrary but uniform ionization fraction. \citet{1994ApJ...426..387H} (hereafter, HF94) extended that analytic expression to include nonuniform ionization fraction over the course of the electrons' transport. We have performed a test to compare how FP performs compared to the E78+HF94 heating rate. For the first step, we configured FP to mimic the assumptions of the CCTTM (labeled FP\_CC\_only).  That is all forces are switched off except Coulomb collisions of nonthermal electrons with ambient electrons and neutral hydrogen. We modeled the transport of electrons injected into CL (Fig.~\ref{fig:loops}) with a power-law flux distribution (Eq.~\ref{eq:powerlaw}) characterized by $E_c = 20$ keV, $\delta = 4$, and an injected energy flux of $\sn{1}{11}$ erg cm$^{-2}$ s$^{-1}$. These are typical values measured during M- and X-class solar flares \citep[e.g.,][]{2003ApJ...595L..97H,2014ApJ...793...70M,2016A&A...588A.115W}. The results are shown in Figure~\ref{fig:fpe78}. The top left panel shows the temperature and density stratification in our loop model. The bottom left panel shows the volumetric heating computed by FP\_CC\_only (black line) compared to the E78+HF94 heating rate (gray dashed line). FP\_CC\_only well reproduces the E78+HF94 analytic model. 

For the second step of this test case, we have switched on the return current, magnetic mirroring, and synchrotron reaction forces (labelled FP). We have injected electrons into the CL loop model plotted in Fig.~\ref{fig:loops} which includes hydrogen and helium ionization fractions. The same power law distribution was injected. The blue line in the bottom left panel of Fig.~\ref{fig:fpe78} shows the resulting volumetric heating rate predicted by FP. More than an order of magnitude more heat is deposited in the coronal portions of the loop than predicted in the E78+HF94 model, since the density is low this can dramatically raise the coronal temperature. The bulk of the coronal portion is due to Joule heating caused by the return current. The peak of the energy deposition is 0.14 Mm higher and primarily in the transition region and top of the chromosphere rather than the deep chromosphere as predicted by the E78+HF94 model. The small bump near 10 Mm is due to the magnetic mirror causing a spreading in pitch angle resulting in a slight increase in column depth and thus more heating. Simulations of flare loops and the resulting emission \citep[e.g.,][]{2005ApJ...630..573A} will behave very differently in response to heating computed using the E78+HF94 model compared to FP. 
 
For the final step of this test case, we use FP to compute the nonthermal bremsstrahlung emitted by the transporting electrons and compare it to the CCTTM prediction calculated from the FP\_CC\_only case. The bremsstrahlung cross section is from \citet{1997A&A...326..417H} as implemented in the SSW IDL routine, \texttt{BRM\_BREMCROSS}. The result is shown in the right panel of Fig.~\ref{fig:fpe78}. As a check to demonstrate the FP\_CC\_only behaves similar to the CCTTM, we also plot the predicted spectrum computed using the  SSW IDL routine, \texttt{BRM2\_NUI}, which implements a nonuniform ionization extension to the CCTTM \citep{2011ApJ...731..106S}. The CCTTM case predicts nearly an order of magnitude more flux than the FP model. This is because the return current force slows the nonthermal electrons, so that much of their energy has been lost by the time they reach the thick target. Clearly, inverting observed X-ray spectra with models that do not account for the return current force will infer very different injected electron distributions compared with models that do include it. To remedy this we have incorporated FP into the X-ray inversion model, OSPEX.    

\subsection{Return Current Model}
\begin{figure*}[htb!]
    \includegraphics[scale=0.65]{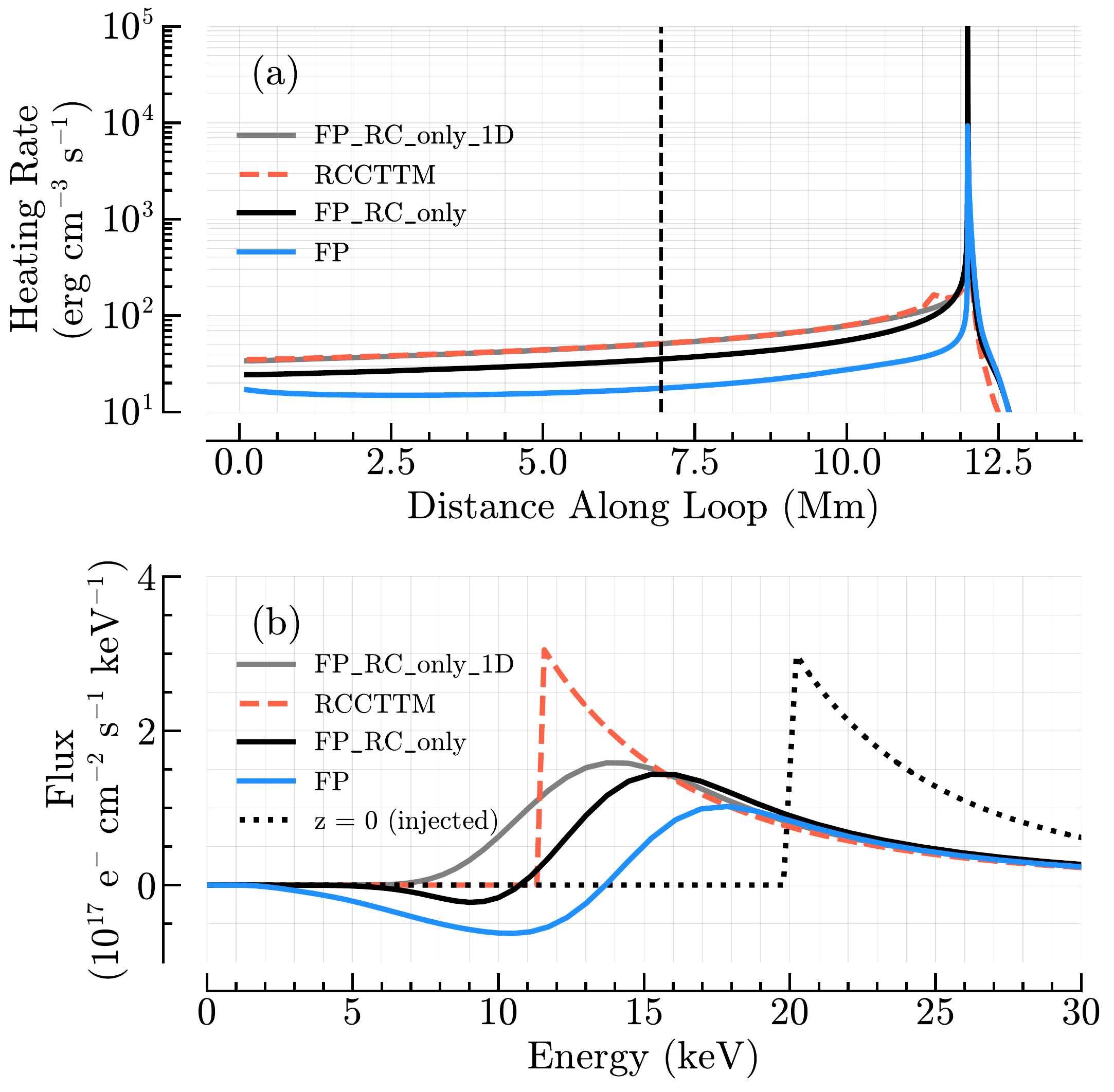}
     \caption{(Top panel) The volumetric heating rates from the FP\_RC\_only\_1D model (black line), FP\_RC\_only (grey line), FP (blue line), and RCCTTM (red line). The vertical dotted line at 7 Mm shows the position of the fluxes plotted in the bottom panel. In each model electrons with a power-law spectrum with $E_c = 20$ keV, $\delta = 4$, and energy flux of $\sn{1}{11}$ erg cm$^{-2}$ s$^{-1}$ were injected into the CL loop model. (Bottom panel) The electron flux distributions at 7 Mm for the same models in the top panel. The dotted black line is the injected flux distribution.}
    \label{fig:compare_fp_h12}
\end{figure*}
For this test case, we demonstrate the performance of FP in conditions where the return current force dominates and compare it to the analytic return current model, RCCTTM developed by \citet{2012ApJ...745...52H}. The plasma resistivity, $\eta$, increases with decreasing temperature and Coulomb collisions decrease with decreasing density, so the return current force dominates in low-temperature, low-density environments.

For comparison with RCCTTM, we put FP into a mode in which it makes similar assumptions, and then, by relaxing those assumptions, we demonstrate their range of applicability. RCCTTM is a 1D model, assuming that all nonthermal electrons have zero pitch angle. The electrons are decelerated by the return current electric field and other forces are neglected by RCCTTM. Once the nonthermal electrons have been reduced in energy to a thermalization energy, $E_{th}$, they are lost from the electron beam. $E_{th}$ is a free parameter of the RCCTTM and is often taken to be $(\delta + 1) k T$ (see Eq.~\ref{eq:dkt}).

We label FP configured to run in a mode that mimics RCCTTM as FP\_RC\_only\_1D. We relax the 1D assumption by allowing the electron pitch angle to change in response to the return current force, but still with all other forces switched off. FP in this mode is labeled, FP\_RC\_only. Finally, FP including all forces described in \S\ref{sec:method} keeps the label, FP. An advantage of FP is that, since it includes the dynamic friction and energy diffusion components of the Coulomb collision force, it is able to directly model thermalization. Therefore, the parameter $E_{th}$ is not needed, and FP naturally produces a Maxwell-Boltzmann distribution as the nonthermal particles undergo many collisions.

As in previous test cases, we model the injection of nonthermal electrons with $E_c = 20$ keV, $\delta = 4$ and energy flux of $\sn{1}{11}$ erg cm$^{-2}$ s$^{-1}$ into the CL loop model. The results of the comparison are shown in Fig.~\ref{fig:compare_fp_h12}. In the top panel, we plot the heating rates obtained using the RCCTTM model (red dashed line) compared to FP\_RC\_only\_1D (gray line), FP\_RC\_only (black line), and FP (blue line). As expected, the heating rates predicted by FP\_RC\_only\_1D and RCCTTM are essentially identical. FP\_RC\_only predicts a smaller heating rate throughout the coronal portion of the loop. Interestingly, FP also predicts a lower heating rate, even though it is including deceleration from Coulomb collisions in addition to return current. A reason for this is that in this low-density loop there is relatively little energy loss from the Coulomb collisions, but the collisions do cause some pitch-angle scattering. This decreases the beam collimation resulting in a smaller current density and return current force.

This effect can be seen in the bottom panel of Fig.~\ref{fig:compare_fp_h12} which compares the electron flux distributions obtained from FP and RCCTTM. In FP\_RC\_only and, even more markedly in FP, the flux at lower energies is directed upward (i.e,. in the negative $\hat{z}$ direction), resulting in a relatively smaller $\mathcal{F}$. For particles moving upward, the return current produces an acceleration. This important effect is not captured in 1D models, thus demonstrating the need for FP's pitch-angle treatment. 
Some numerical diffusion across the sharp low-energy cutoff is evident in the FP\_RC\_only\_1D case. Increasing the energy resolution can reduce that effect. We have compared the predicted heating rates and bremsstrahlung spectra from FP\_RC\_only\_1D and RCCTTM and found them to be nearly identical, so we conclude that the current resolution is sufficient.
\subsection{Magnetic Mirroring}\label{sec:magmirror}
\begin{figure*}[htb!]
\includegraphics[scale=.65]{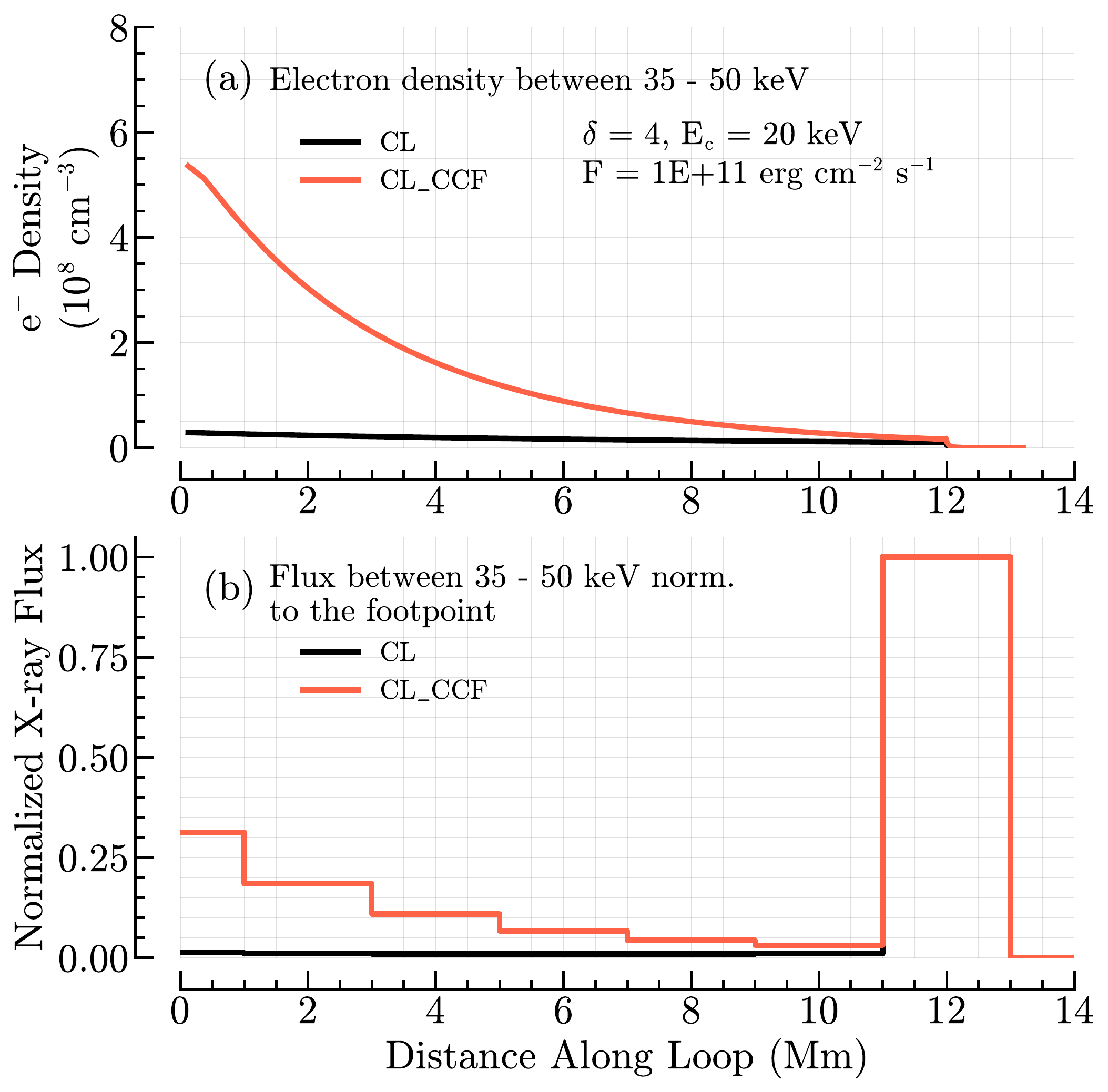}
\caption{(Top panel) The density of nonthermal electrons with energies between $35 - 50$ keV as a function of distance from looptop injected into the CL (black line) and CL\_CCF (red line) loop models. In each case a power-law spectrum characterized by $E_c = 20$ keV, $\delta = 4$, and energy flux of $\sn{1}{11}$ erg cm$^{-2}$ s$^{-1}$ is injected. (Bottom panel) The X-ray bremsstrahlung flux between $35 - 50$ keV integrated over 2 Mm distance intervals  resulting from the nonthermal electron densities shown in the top panel. The X-ray flux has been normalized to the value in the footpoint interval ($11 - 13$ Mm).}\label{fig:magmirror}
\end{figure*}
During flares, nonthermal hard X-rays are occasionally seen near looptops in addition to the footpoints \citep[e.g.,][]{2018ApJ...867...82D}. A likely explanation is that a strongly converging coronal magnetic field forms a magnetic mirror that traps a fraction of the injected particles and producing a coronal X-ray source \citep{2019ApJ...887L..37K}. In this test case, we use FP to demonstrate this trapping effect. We use a flux of nonthermal electrons similar to the previous test cases. That is a power-law with $E_c = 20$~keV, $\delta = 4$, and energy flux of $10^{11}$ erg cm$^{-2}$ s$^{-1}$. We use fluxes with isotropic pitch-angle distributions, so as to increase the effect of magnetic mirroring, which is strongest for particles with high pitch-angle. We use FP to model the injection of this flux into the CL and CL\_CCF loop models, which have uniform and strongly converging coronal fields, respectively (Fig.~\ref{fig:loops}).

The results of this test case are shown in Fig.~\ref{fig:magmirror}. The top panel shows the nonthermal electron density for electrons with energies between $35 - 50$~keV as a function of distance from looptop. In the bottom panel, we show the bremsstrahlung X-ray flux also in the $35 - 50$~keV band, integrated over 2 Mm distance intervals, and normalized to the flux in the footpoint interval ($11 - 13$ Mm). As in \S\ref{sec:ccttm} the bremsstrahlung cross section is from \citet{1997A&A...326..417H}. In each panel the black and red lines are the results from injections into the CL and CL\_CCF loop models, respectively. In the CL model, which has uniform coronal magnetic field, there is little emission outside of the footpoint. In CL\_CCF, the magnetic mirroring force is sufficiently strong to trap electrons near the looptop producing a nonthermal looptop X-ray source that is more than 30\% as bright as the footpoint. 
\subsection{Energy Diffusion}\label{sec:ed}
\begin{figure*}[htb!]
\begin{interactive}{animation}{FP_PlotED_GSK.mp4}
\includegraphics[scale=0.65]{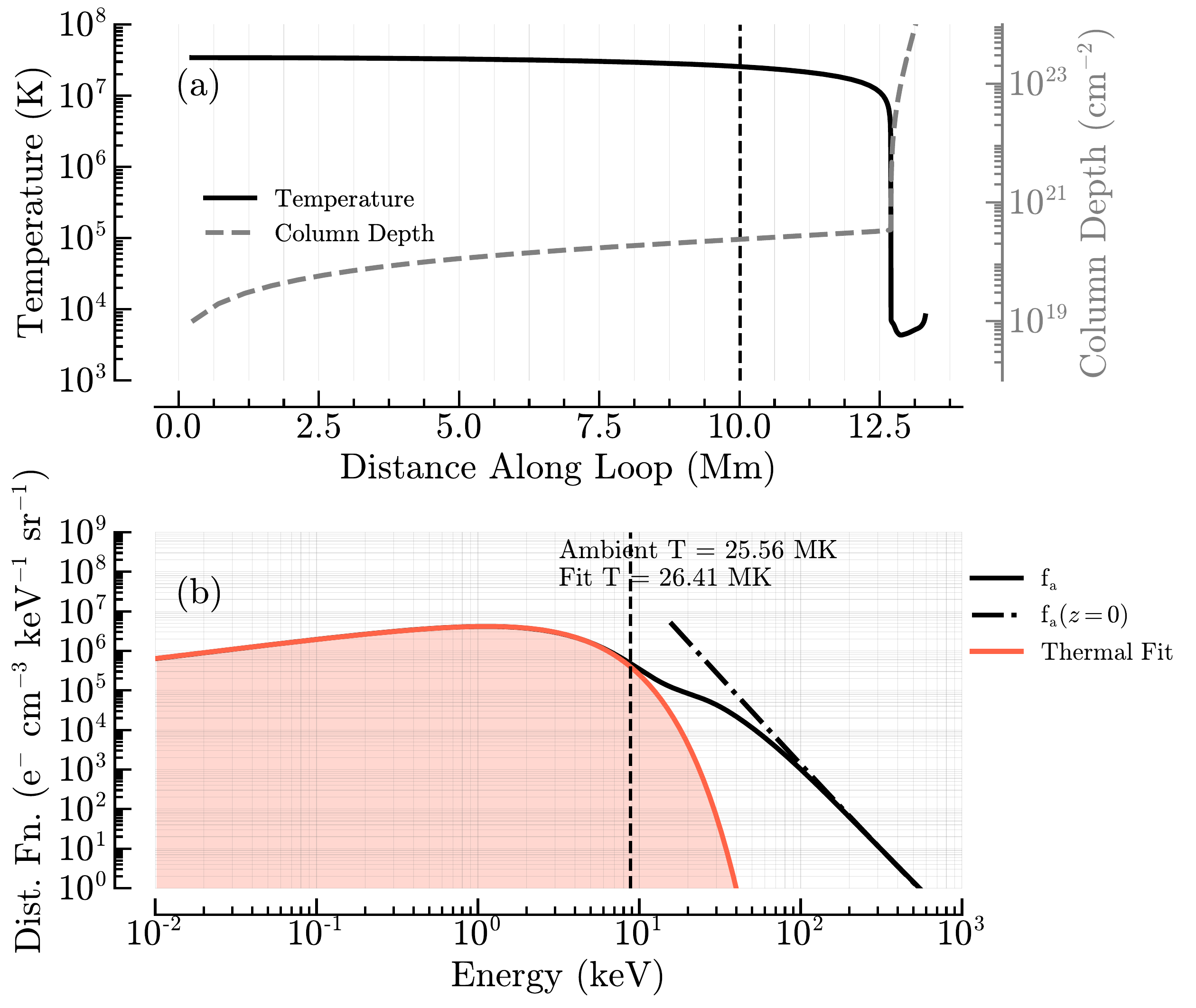}
\end{interactive}
\caption{(Top panel) Temperature (solid black line) and column depth (dashed grey line) in the HL loop model. The vertical dashed line indicates the $z$ position of the distribution function plotted in the lower panel. (Bottom panel) Distribution function, $\fa(E,\mu = 0, z)$ (solid black line) for nonthermal electrons injected into the HL loop model, at the position indicated in panel (a).  For comparison the injected distribution function, $\fa(E,\mu = 0, z = 0)$, is also plotted (dash-dotted black line). The lower energy portion of $\fa$ was fitted to a Maxwell-Boltzmann function (red line and shaded area). The temperature of the fit and ambient electron temperature are stated on the panel. The energy, $(\delta+1)k T$, is indicated by the dashed vertical black line.This figure is available as an animation. The animation follows the sequence from an ambient temperature of 34.14 MK to 0.10 MK. The realtime duration of the video is 22 seconds.
}\label{fig:ed}
\end{figure*}

\begin{figure*}[htb!]
\includegraphics[scale=0.65]{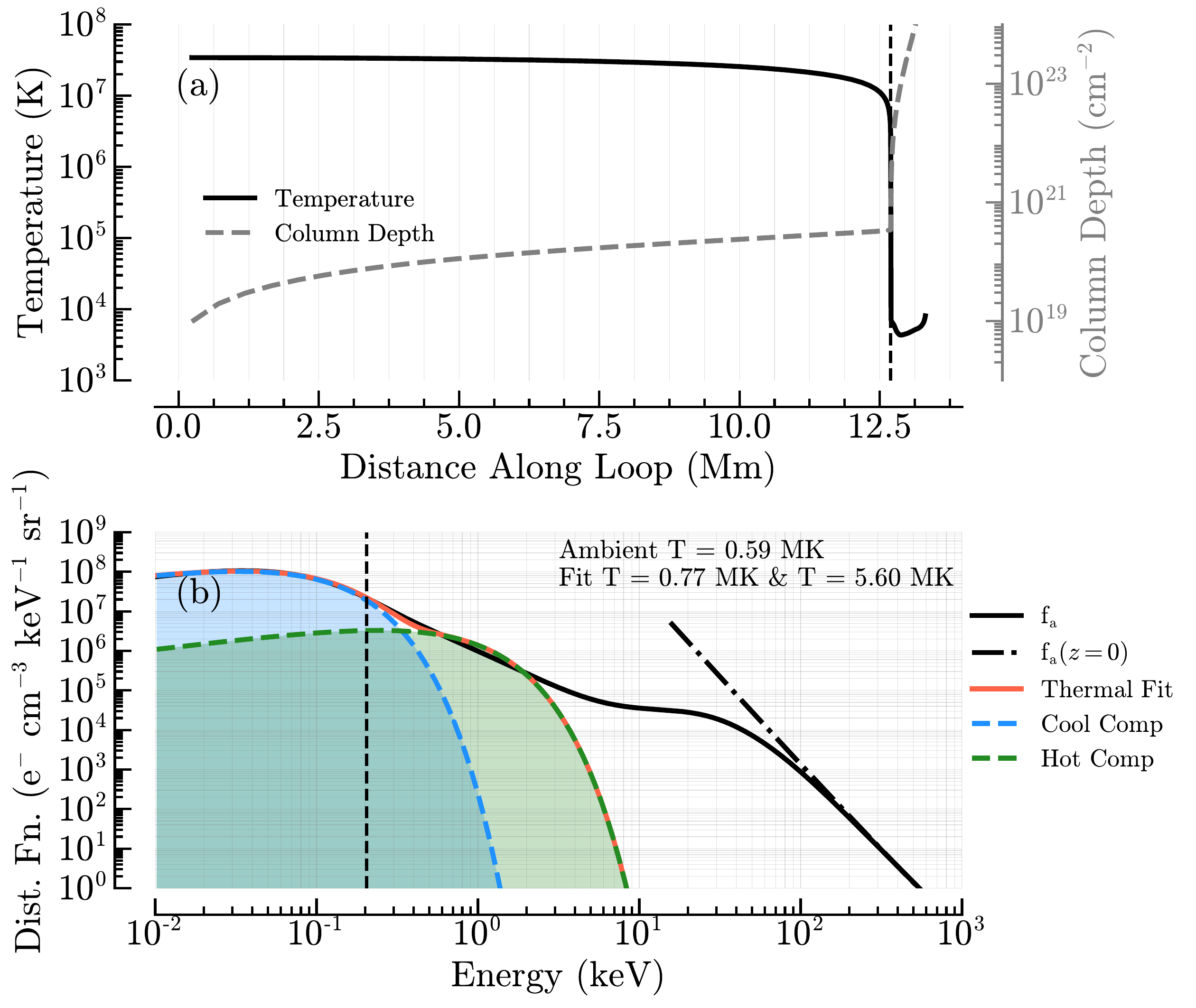}
\caption{Same as Figure~\ref{fig:ed} but for a different position along the loop. In the transition region, the low energy portion of $\fa$ is best fit by two Maxwell-Boltzmann functions. Their sum, the hotter component and the cooler component are indicated by the solid red, dashed blue, and dashed green lines, respectively.}\label{fig:ed2}
\end{figure*}

For this test case, we consider the importance of parallel momentum (i.e., energy) diffusion. This is the $\hat{p}$ component of $D_c$ (Eq.~\ref{eq:fcdc}), and gives rise to the second derivative, $\partial^2 \fa/\partial p^2$, in Eq.~\ref{eqn:ssfp}. To understand when energy diffusion is important relative to dynamic friction, we take the ratio,
\begin{equation}\label{eq:dkt}
\frac{D_c}{F_c} = \frac{p \pd{\fa}{p}}{2 x_e \fa} = \frac{-(\delta +1) k T}{E}
\end{equation} 
where we have assumed that the flux is a power-law with respect to energy (i.e., has a form like Eq.~\ref{eq:powerlaw}), so $\fa$ is a power-law with index, $-2 (\delta + 1)$, with respect to momentum. Here we consider the collision of nonthermal electrons on thermal electron targets, so that the mass ratio appearing in $F_c$ cancels. In these circumstances, $D_c$ becomes dominant for energies less than $(\delta+1) k T$. This is closely related to the warm-target approximation discussed in the proton beam test case (\S\ref{sec:protonbeam}).  Flares often heat a large emission measure of plasma to greater than 30 MK. $(\delta+1) k T = 13$~keV for $\delta = 4$ and $T = 30$ MK, which is comparable to cutoff energies deduced from the transition energy between the thermal to nonthermal components of X-ray spectra in many flares. $D_c$ is in the positive $\hat{p}$ direction, meaning that these low energy electrons experience a net acceleration, distorting $\fa$ away from a power-law. The steady state is achieved when $D_c$ balances $F_c$, in which case $\fa$ is a Maxwell-Boltzmann distribution and the electrons are thermalized. 

To demonstrate this effect we consider the injection of a power-law distribution of nonthermal electrons into our hot loop model, HL, which has an apex temperature of 34~MK. We choose the relatively low value of $E_c = 15$~keV to highlight the effects of energy diffusion. As in previous cases, we choose $\delta = 4$ and an energy flux of $\sn{1}{11}$ erg cm$^{-2}$ s$^{-1}$. At 34 MK, $(\delta+1) k T = 15$~keV. The results are shown in Fig.~\ref{fig:ed}. In this loop, the coronal density is high ($>10^{11}$ cm$^{-3}$) and the column depth (gray dashed line in the top panel) quickly becomes large. Above 1.25~Mm, $\fa$ (black line in the bottom panel) begins to resemble a Maxwell-Boltzmann distribution (red line in the bottom panel)\textemdash albeit much hotter than the ambient temperature\textemdash for energies less than 5 keV. The column depth continues to increase as the electrons move down the loop, and by 9~Mm, electrons with $E < 10$ keV are essentially fully thermalized. The lower energy side of $\fa$ is well fit by a Maxwell-Boltzmann distribution with temperature comparable to the ambient electron temperature. In the transition region $\fa$ is best fit at low energy by two Maxwell-Boltzmann distributions as shown in Figure~\ref{fig:ed2}. The cooler distribution (dashed green line in the bottom panel) has a temperature comparable to that of the local ambient plasma and the hotter distribution (dashed blue line in the bottom panel) has $T\sim 10$ MK and represents the heat flux carried by the injected electrons moving through the transition region. The vertical black dashed line in the bottom panel indicates the energy, $E = (\delta+1)k T$. This is a good indicator for the energy above which $\fa$ deviates from a thermal distribution.

\subsection{Proton Beams}\label{sec:protonbeam}
\begin{figure*}[htb!]
\includegraphics[scale=0.65]{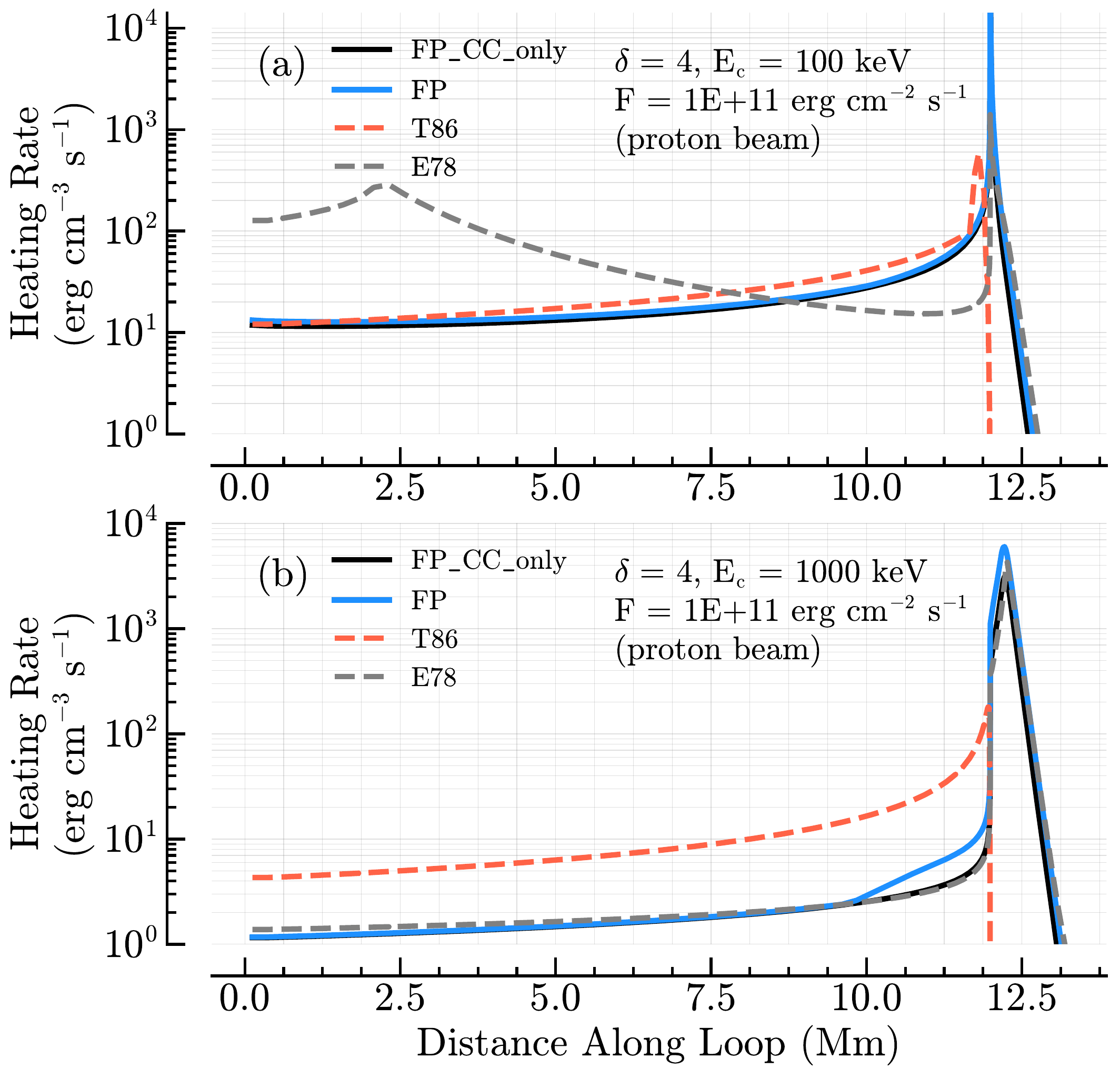}
\caption{(Top panel) The heating rate for a power-law distribution of protons with $E_c = 100$ keV, $\delta = 4$ and injected energy flux of $\sn{1}{11}$ erg cm$^{-2}$ s$^{-1}$ injected into the CL loop model computed using FP (blue), FP\_CC\_only (black), \citet{1986ApJ...309..409T} warm-target (dashed red), and E78 cold-target (dashed grey). (Bottom panel) Same as above except for a power-law distribution with $E_c = 1$ MeV.}\label{fig:protons}
\end{figure*}
In this test case, we demonstrate how FP models the transport of nonthermal protons injected into the CL loop model (Fig.~\ref{fig:loops}). We consider the injection of protons with a power-law energy distribution for two different cutoff energies, 100~keV and 1 MeV. These values were chosen to demonstrate the warm- and cold-target regimes. The results are shown in the top and bottom panels of Fig.~\ref{fig:protons}, respectively. In the CL loop model the corona is at a temperature of 3.4~MK so the speed of 100~keV protons is slower than the thermal electron speed but faster than the thermal proton speed. That is $x_e = 0.19$ and $x_p = 344$ (Eq.~\ref{eqn:xi}). This is the ``warm-target" regime as defined by \citet{1986ApJ...309..409T}. In that work they obtained an analytic expression (their Eq.~38) for the heating rate produced by the injection of proton beams on a warm-target.  We have plotted their analytic expression (T86; red dashed line) in Fig.~\ref{fig:protons} and compared it to the result from E78 for cold-target collisions (gray dashed line), and the heating rates predicted by FP with only the Coulomb collision force on (FP\_CC\_only; black line) and FP with all forces on (blue line). We emphasize that both FP and FP\_CC\_only do \emph{not} make the warm- or cold-target assumptions but rather use a full form of the Coloumb collision operator and are applicable in both extremes and between them. This is important for modeling the transport of particles as they move from one regime into the other. For example, in the top panel of Fig.~\ref{fig:protons}, the 100 keV protons are in the warm-target regime as they transport through the corona but as they move into the transition region, the cold-target regime becomes more applicable. FP smoothly captures the collisions in both regions. Interestingly, since the heating rates predicted by FP and T86 are much less than that of E78 in the corona, it is evident that collisions in the warm-target regime are much less effective at slowing particles than predicted by cold-target theory. This is important in predicting to what column depth particles will penetrate.

For the case of 1~MeV protons, $x_e = 1.9$ in the corona and the cold-target theory is more applicable throughout the protons' transport. This is shown in the bottom panel of Figure~\ref{fig:protons}. The cold-target heating rate (gray dashed line) closely matches the FP\_CC\_only rate throughout the entire loop. 

Interestingly, since the heating rates in the FP\_CC\_only and FP cases are similar, it is evident that the return current has only a small effect on these nonthermal protons. The return current causes a potential drop of a few keV but that is relatively small compared to their injected energies. The increased heating rate between 10 -- 12 Mm in the FP compared to FP\_CC\_only cases is due to the magnetic mirror causing an increase in pitch-angle and thus an increase in column depth. 
\section{Conclusions} \label{sec:conc}
We have developed a method, called FP, to model the transport of nonthermal particles injected at the top of magnetic loops in stellar atmospheres. The nonthermal particles are widely believed to be a dominant source of energy transport in flares, so accurately modeling their transport is critical for understanding flare energetics. Their interactions with the ambient atmosphere are responsible for producing the ubiquitous hard X-ray bremsstrahlung and radio synchrotron (in the case of nonthermal electrons) and $\gamma$-ray line emission (in the case of nonthermal protons). By inverting hard X-ray, radio and $\gamma$-ray observations during flares, details of the injected particles can be inferred. By accounting for numerous transport effects, FP can be a powerful tool for performing these inversions. We have incorporated FP into the X-ray data analysis tool, OSPEX. In a forthcoming study, we will describe using FP combined with OSPEX to determine injected electron distributions during solar flares.    

Our method solves the steady state Fokker-Plank equation accounting for the effects of Coulomb collisions including dynamic friction and energy diffusion, return currents, magnetic mirroring, and synchrotron emission. These are the dominant forces acting on the particles during their transport. We have demonstrated the performance of FP using several test cases. We showed that in cool loops, the return current can be a dominant force on nonthermal electrons. The return current force significantly alters the location of heating compared to models which neglect the return current. We have used FP to show an example of a strongly converging magnetic field in the corona trapping energetic electrons and producing a looptop nonthermal X-ray source. We have demonstrated when energy diffusion can be significant in comparison to the dynamic friction, and showed how including both allows our model to capture the thermalization process. We have demonstrated that for $\sim 100$ keV protons, warm-target collisions are much less effective at slowing the protons than predicted from cold-target theory. This allows them to penetrate deeper. For protons, we have shown that return current is not a significant effect, since the return current causes a potential drop that is a very small fraction of their kinetic energy.   

We have taken care to make FP as accurate as possible. However, there are a few limitations to our method, listed below.
\begin{enumerate}
 \item Our method neglects ``self-collisions.'' That is we neglect the effect of nonthermal particles colliding with other nonthermal particles in the same distribution, $\fa$. We expect this effect to be small because the number density of nonthermal particles is typically much less than that of the ambient plasma. But in future work we plan to remedy this limitation. The method is straightforward. When iteratively solving Eq.~\ref{eqn:mateqn}, the current solution, $\fa^n$, is used to evaluate the Rosenbluth potentials (Eq.~\ref{eqn:potentials}), and the corresponding forces are calculated using Eq.~\ref{eqn:fd} and added to the net collision force. 
 \item Our model of the return current does not include a component of ambient electrons accelerated by the return current electric field into the runaway regime \citep{1985ApJ...293..584H}. We are currently developing a model to include runaways and in future work we will incorporate that into FP.
 \item FP does not include beam and current driven instabilities so is not applicable in regimes where those instabilities become dominant.
 \item FP models ``half-loops'' from looptop to footpoint with the assumption that a full loop is symmetric about the looptop. Thus, FP is not able to model full loops with asymmetric legs.
\end{enumerate}

We have demonstrated that FP predicts significantly different heating rates than the analytical approach (E78 \& E78+HF94) and the Coulomb-collision-only (CC\_only) approaches that are commonly used in solar/stellar flare modelling. FP is currently being merged with the flare radiation hydrodynamic model RADYN. Using this we will investigate, in a future work, the resulting impact of this more physically accurate heating rate on flare dynamics.

\acknowledgments
JCA acknowledges funding from NASA's Heliophysics Innovation Fund and NASA's Heliophysics Supporting Research program. MA was supported by NASA's Heliophysics Supporting Research fund. AFK acknowledges funding from NSF AGS Solar-Terrestrial Award Number 1916511, and he thanks Prof. Lyndsay Fletcher for helpful discussions about velocity diffusion in solar flares. GSK was funded by an appointment to the NASA Postdoctoral Program at Goddard Space Flight Center, administered by USRA through a contract with NASA, and by the Heliophysics Innovation Fund. This research benefited from discussions held at a meeting of Graham Kerr and Vanessa Polito’s International Space Science Institute team: `Interrogating Field-Aligned Solar Flare Models: Comparing, Contrasting and Improving.’

\bibliography{fp_paper}

\begin{thebibliography}{}
\expandafter\ifx\csname natexlab\endcsname\relax\def\natexlab#1{#1}\fi
\providecommand{\url}[1]{\href{#1}{#1}}

\bibitem[{{Abbett} \& {Hawley}(1999)}]{1999ApJ...521..906A}
{Abbett}, W.~P., \& {Hawley}, S.~L. 1999, \apj, 521, 906

\bibitem[{{Alaoui} \& {Holman}(2017)}]{2017ApJ...851...78A}
{Alaoui}, M., \& {Holman}, G.~D. 2017, \apj, 851, 78

\bibitem[{{Alaoui} {et~al.}(2019){Alaoui}, {Krucker}, \&
  {Saint-Hilaire}}]{2019SoPh..294..105A}
{Alaoui}, M., {Krucker}, S., \& {Saint-Hilaire}, P. 2019, \solphys, 294, 105

\bibitem[{{Allred} {et~al.}(2005){Allred}, {Hawley}, {Abbett}, \&
  {Carlsson}}]{2005ApJ...630..573A}
{Allred}, J.~C., {Hawley}, S.~L., {Abbett}, W.~P., \& {Carlsson}, M. 2005,
  \apj, 630, 573

\bibitem[{{Allred} {et~al.}(2015){Allred}, {Kowalski}, \&
  {Carlsson}}]{2015ApJ...809..104A}
{Allred}, J.~C., {Kowalski}, A.~F., \& {Carlsson}, M. 2015, \apj, 809, 104

\bibitem[{{Aschwanden}(2002)}]{2002SSRv..101....1A}
{Aschwanden}, M.~J. 2002, \ssr, 101, 1

\bibitem[{{Aschwanden} {et~al.}(2016){Aschwanden}, {Holman}, {O'Flannagain},
  {Caspi}, {McTiernan}, \& {Kontar}}]{2016ApJ...832...27A}
{Aschwanden}, M.~J., {Holman}, G., {O'Flannagain}, A., {et~al.} 2016, \apj,
  832, 27

\bibitem[{{Battaglia} \& {Benz}(2008)}]{2008A&A...487..337B}
{Battaglia}, M., \& {Benz}, A.~O. 2008, \aap, 487, 337

\bibitem[{{Battaglia} {et~al.}(2012){Battaglia}, {Kontar}, {Fletcher}, \&
  {MacKinnon}}]{2012ApJ...752....4B}
{Battaglia}, M., {Kontar}, E.~P., {Fletcher}, L., \& {MacKinnon}, A.~L. 2012,
  \apj, 752, 4

\bibitem[{{Benz}(2002)}]{2002ASSL..279.....B}
{Benz}, A. 2002, {Plasma Astrophysics, second edition}, Vol. 279,
  doi:10.1007/978-0-306-47719-5

\bibitem[{{Brown}(1971)}]{1971SoPh...18..489B}
{Brown}, J.~C. 1971, \solphys, 18, 489

\bibitem[{{Carlsson} \& {Stein}(1992)}]{1992ApJ...397L..59C}
{Carlsson}, M., \& {Stein}, R.~F. 1992, \apjl, 397, L59

\bibitem[{{Carlsson} \& {Stein}(1995)}]{1995ApJ...440L..29C}
---. 1995, \apjl, 440, L29

\bibitem[{{Carlsson} \& {Stein}(1997)}]{1997ApJ...481..500C}
---. 1997, \apj, 481, 500

\bibitem[{{Caspi} {et~al.}(2014){Caspi}, {Krucker}, \&
  {Lin}}]{2014ApJ...781...43C}
{Caspi}, A., {Krucker}, S., \& {Lin}, R.~P. 2014, \apj, 781, 43

\bibitem[{{Chandrasekhar}(1943)}]{1943RvMP...15....1C}
{Chandrasekhar}, S. 1943, Reviews of Modern Physics, 15, 1

\bibitem[{{Dennis} {et~al.}(2018){Dennis}, {Duval-Poo}, {Piana}, {Inglis},
  {Emslie}, {Guo}, \& {Xu}}]{2018ApJ...867...82D}
{Dennis}, B.~R., {Duval-Poo}, M.~A., {Piana}, M., {et~al.} 2018, \apj, 867, 82

\bibitem[{{Dreicer}(1960)}]{1960PhRv..117..329D}
{Dreicer}, H. 1960, Physical Review, 117, 329

\bibitem[{{Emslie}(1978)}]{1978ApJ...224..241E}
{Emslie}, A.~G. 1978, \apj, 224, 241

\bibitem[{{Emslie}(1980)}]{1980ApJ...235.1055E}
---. 1980, \apj, 235, 1055

\bibitem[{{Emslie} {et~al.}(2018){Emslie}, {Bian}, \&
  {Kontar}}]{2018ApJ...862..158E}
{Emslie}, A.~G., {Bian}, N.~H., \& {Kontar}, E.~P. 2018, \apj, 862, 158

\bibitem[{{Emslie} {et~al.}(1998){Emslie}, {Mariska}, {Montgomery}, \&
  {Newton}}]{1998ApJ...498..441E}
{Emslie}, A.~G., {Mariska}, J.~T., {Montgomery}, M.~M., \& {Newton}, E.~K.
  1998, \apj, 498, 441

\bibitem[{{Emslie} {et~al.}(2012){Emslie}, {Dennis}, {Shih}, {Chamberlin},
  {Mewaldt}, {Moore}, {Share}, {Vourlidas}, \& {Welsch}}]{2012ApJ...759...71E}
{Emslie}, A.~G., {Dennis}, B.~R., {Shih}, A.~Y., {et~al.} 2012, \apj, 759, 71

\bibitem[{{Evans}(1955)}]{evans55}
{Evans}, R.~D. 1955, {The Atomic Nucleus} (New York: McGraw-Hill)

\bibitem[{{Fisher} {et~al.}(1985){Fisher}, {Canfield}, \&
  {McClymont}}]{1985ApJ...289..414F}
{Fisher}, G.~H., {Canfield}, R.~C., \& {McClymont}, A.~N. 1985, \apj, 289, 414

\bibitem[{{Gordovskyy} {et~al.}(2005){Gordovskyy}, {Zharkova}, {Voitenko}, \&
  {Goossens}}]{2005AdSpR..35.1743G}
{Gordovskyy}, M., {Zharkova}, V.~V., {Voitenko}, Y.~M., \& {Goossens}, M. 2005,
  Advances in Space Research, 35, 1743

\bibitem[{{Haug}(1997)}]{1997A&A...326..417H}
{Haug}, E. 1997, \aap, 326, 417

\bibitem[{{Hawley} \& {Fisher}(1994)}]{1994ApJ...426..387H}
{Hawley}, S.~L., \& {Fisher}, G.~H. 1994, \apj, 426, 387

\bibitem[{{Hirshman}(1977)}]{1977PhFl...20..589H}
{Hirshman}, S.~P. 1977, Physics of Fluids, 20, 589

\bibitem[{{Holman}(1985)}]{1985ApJ...293..584H}
{Holman}, G.~D. 1985, \apj, 293, 584

\bibitem[{{Holman}(2012)}]{2012ApJ...745...52H}
---. 2012, \apj, 745, 52

\bibitem[{{Holman} {et~al.}(2003){Holman}, {Sui}, {Schwartz}, \&
  {Emslie}}]{2003ApJ...595L..97H}
{Holman}, G.~D., {Sui}, L., {Schwartz}, R.~A., \& {Emslie}, A.~G. 2003, \apjl,
  595, L97

\bibitem[{{Hurford} {et~al.}(2006){Hurford}, {Krucker}, {Lin}, {Schwartz},
  {Share}, \& {Smith}}]{2006ApJ...644L..93H}
{Hurford}, G.~J., {Krucker}, S., {Lin}, R.~P., {et~al.} 2006, \apjl, 644, L93

\bibitem[{{Jeffrey} {et~al.}(2019){Jeffrey}, {Kontar}, \&
  {Fletcher}}]{2019ApJ...880..136J}
{Jeffrey}, N. L.~S., {Kontar}, E.~P., \& {Fletcher}, L. 2019, \apj, 880, 136

\bibitem[{{Karlick{\'y}}(2009)}]{2009ApJ...690..189K}
{Karlick{\'y}}, M. 2009, \apj, 690, 189

\bibitem[{{Ka{\v{s}}parov{\'a}} {et~al.}(2009){Ka{\v{s}}parov{\'a}}, {Varady},
  {Heinzel}, {Karlick{\'y}}, \& {Moravec}}]{2009A&A...499..923K}
{Ka{\v{s}}parov{\'a}}, J., {Varady}, M., {Heinzel}, P., {Karlick{\'y}}, M., \&
  {Moravec}, Z. 2009, \aap, 499, 923

\bibitem[{{Knight} \& {Sturrock}(1977)}]{1977ApJ...218..306K}
{Knight}, J.~W., \& {Sturrock}, P.~A. 1977, \apj, 218, 306

\bibitem[{{Kong} {et~al.}(2019){Kong}, {Guo}, {Shen}, {Chen}, {Chen}, {Musset},
  {Glesener}, {Pongkitiwanichakul}, \& {Giacalone}}]{2019ApJ...887L..37K}
{Kong}, X., {Guo}, F., {Shen}, C., {et~al.} 2019, \apjl, 887, L37

\bibitem[{{Kontar} {et~al.}(2011){Kontar}, {Brown}, {Emslie}, {Hajdas},
  {Holman}, {Hurford}, {Ka{\v{s}}parov{\'a}}, {Mallik}, {Massone}, {McConnell},
  {Piana}, {Prato}, {Schmahl}, \& {Suarez-Garcia}}]{2011SSRv..159..301K}
{Kontar}, E.~P., {Brown}, J.~C., {Emslie}, A.~G., {et~al.} 2011, \ssr, 159, 301

\bibitem[{{Kowalski} {et~al.}(2015){Kowalski}, {Hawley}, {Carlsson}, {Allred},
  {Uitenbroek}, {Osten}, \& {Holman}}]{2015SoPh..290.3487K}
{Kowalski}, A.~F., {Hawley}, S.~L., {Carlsson}, M., {et~al.} 2015, \solphys,
  290, 3487

\bibitem[{{Leach} \& {Petrosian}(1981)}]{1981ApJ...251..781L}
{Leach}, J., \& {Petrosian}, V. 1981, \apj, 251, 781

\bibitem[{{Lin} {et~al.}(2002){Lin}, {Dennis}, {Hurford}, {Smith}, {Zehnder},
  {Harvey}, {Curtis}, {Pankow}, {Turin}, {Bester}, {Csillaghy}, {Lewis},
  {Madden}, {van Beek}, {Appleby}, {Raudorf}, {McTiernan}, {Ramaty}, {Schmahl},
  {Schwartz}, {Krucker}, {Abiad}, {Quinn}, {Berg}, {Hashii}, {Sterling},
  {Jackson}, {Pratt}, {Campbell}, {Malone}, {Landis}, {Barrington-Leigh},
  {Slassi-Sennou}, {Cork}, {Clark}, {Amato}, {Orwig}, {Boyle}, {Banks},
  {Shirey}, {Tolbert}, {Zarro}, {Snow}, {Thomsen}, {Henneck}, {McHedlishvili},
  {Ming}, {Fivian}, {Jordan}, {Wanner}, {Crubb}, {Preble}, {Matranga}, {Benz},
  {Hudson}, {Canfield}, {Holman}, {Crannell}, {Kosugi}, {Emslie}, {Vilmer},
  {Brown}, {Johns-Krull}, {Aschwanden}, {Metcalf}, \&
  {Conway}}]{2002SoPh..210....3L}
{Lin}, R.~P., {Dennis}, B.~R., {Hurford}, G.~J., {et~al.} 2002, \solphys, 210,
  3

\bibitem[{{Liu} {et~al.}(2009){Liu}, {Petrosian}, \&
  {Mariska}}]{2009ApJ...702.1553L}
{Liu}, W., {Petrosian}, V., \& {Mariska}, J.~T. 2009, \apj, 702, 1553

\bibitem[{{MacKinnon} \& {Craig}(1991)}]{1991A&A...251..693M}
{MacKinnon}, A.~L., \& {Craig}, I.~J.~D. 1991, \aap, 251, 693

\bibitem[{{Mart{\'\i}nez-Sykora} {et~al.}(2012){Mart{\'\i}nez-Sykora}, {De
  Pontieu}, \& {Hansteen}}]{2012ApJ...753..161M}
{Mart{\'\i}nez-Sykora}, J., {De Pontieu}, B., \& {Hansteen}, V. 2012, \apj,
  753, 161

\bibitem[{{Mauas} \& {G{\'o}mez}(1997)}]{1997ApJ...483..496M}
{Mauas}, P. J.~D., \& {G{\'o}mez}, D.~O. 1997, \apj, 483, 496

\bibitem[{{McTiernan} \& {Petrosian}(1990)}]{1990ApJ...359..524M}
{McTiernan}, J.~M., \& {Petrosian}, V. 1990, \apj, 359, 524

\bibitem[{{Milligan} {et~al.}(2014){Milligan}, {Kerr}, {Dennis}, {Hudson},
  {Fletcher}, {Allred}, {Chamberlin}, {Ireland}, {Mathioudakis}, \&
  {Keenan}}]{2014ApJ...793...70M}
{Milligan}, R.~O., {Kerr}, G.~S., {Dennis}, B.~R., {et~al.} 2014, \apj, 793, 70

\bibitem[{{Papadopoulos}(1977)}]{1977RvGSP..15..113P}
{Papadopoulos}, K. 1977, Reviews of Geophysics and Space Physics, 15, 113

\bibitem[{{Park} \& {Petrosian}(1996)}]{1996ApJS..103..255P}
{Park}, B.~T., \& {Petrosian}, V. 1996, \apjs, 103, 255

\bibitem[{{Petkaki} {et~al.}(2012){Petkaki}, {Del Zanna}, {Mason}, \&
  {Bradshaw}}]{2012A&A...547A..25P}
{Petkaki}, P., {Del Zanna}, G., {Mason}, H.~E., \& {Bradshaw}, S.~J. 2012,
  \aap, 547, A25

\bibitem[{{Rosenbluth} {et~al.}(1957){Rosenbluth}, {MacDonald}, \&
  {Judd}}]{1957PhRv..107....1R}
{Rosenbluth}, M.~N., {MacDonald}, W.~M., \& {Judd}, D.~L. 1957, Physical
  Review, 107, 1

\bibitem[{{Rowland} \& {Vlahos}(1985)}]{1985A&A...142..219R}
{Rowland}, H.~L., \& {Vlahos}, L. 1985, \aap, 142, 219

\bibitem[{{Rubio da Costa} {et~al.}(2016){Rubio da Costa}, {Kleint},
  {Petrosian}, {Liu}, \& {Allred}}]{2016ApJ...827...38R}
{Rubio da Costa}, F., {Kleint}, L., {Petrosian}, V., {Liu}, W., \& {Allred},
  J.~C. 2016, \apj, 827, 38

\bibitem[{{Sim{\~o}es} \& {Kontar}(2013)}]{2013A&A...551A.135S}
{Sim{\~o}es}, P.~J.~A., \& {Kontar}, E.~P. 2013, \aap, 551, A135

\bibitem[{{Snyder} \& {Scott}(1949)}]{1949PhRv...76..220S}
{Snyder}, H.~S., \& {Scott}, W.~T. 1949, Physical Review, 76, 220

\bibitem[{{Spitzer}(1962)}]{1962pfig.book.....S}
{Spitzer}, L. 1962, {Physics of Fully Ionized Gases}

\bibitem[{{Su} {et~al.}(2011){Su}, {Holman}, \& {Dennis}}]{2011ApJ...731..106S}
{Su}, Y., {Holman}, G.~D., \& {Dennis}, B.~R. 2011, \apj, 731, 106

\bibitem[{{Tamres} {et~al.}(1986){Tamres}, {Canfield}, \&
  {McClymont}}]{1986ApJ...309..409T}
{Tamres}, D.~H., {Canfield}, R.~C., \& {McClymont}, A.~N. 1986, \apj, 309, 409

\bibitem[{{Trubnikov}(1965)}]{1965RvPP....1..105T}
{Trubnikov}, B.~A. 1965, Reviews of Plasma Physics, 1, 105

\bibitem[{{van den Oord}(1990)}]{1990A&A...234..496V}
{van den Oord}, G.~H.~J. 1990, \aap, 234, 496

\bibitem[{{Warmuth} \& {Mann}(2016)}]{2016A&A...588A.115W}
{Warmuth}, A., \& {Mann}, G. 2016, \aap, 588, A115

\bibitem[{{Zharkova} \& {Gordovskyy}(2005)}]{2005A&A...432.1033Z}
{Zharkova}, V.~V., \& {Gordovskyy}, M. 2005, \aap, 432, 1033

\bibitem[{{Zharkova} \& {Gordovskyy}(2006)}]{2006ApJ...651..553Z}
---. 2006, \apj, 651, 553

\end{thebibliography}

\end{document}